\documentclass[twocolumn,preprintnumbers, amssymb,amsmath,aps,floatfix,prd,nofootinbib,superscriptaddress,showpacs,10pt]{revtex4-1}

\usepackage{hyperref}
\usepackage[all]{hypcap}

\usepackage{graphicx}
\usepackage{epstopdf}
\usepackage{amsmath,amsthm,amssymb}

\usepackage{color,xcolor}
\usepackage{ulem}

\begin{document}

\title{$\gamma$-hadron spectra in p + Pb collisions at $\sqrt{s_{\rm NN}}=5.02$ TeV}

\author{Man Xie}
\affiliation{Key Laboratory of Quark and Lepton Physics (MOE) and Institute
of Particle Physics, Central China Normal University, Wuhan 430079, China}

\author{Xin-Nian Wang}
\affiliation{Key Laboratory of Quark and Lepton Physics (MOE) and Institute
of Particle Physics, Central China Normal University, Wuhan 430079, China}
\affiliation{Nuclear Science Division, Lawrence Berkeley National Laboratory, Berkeley, CA 94720, USA}

\author{Han-Zhong Zhang}
\affiliation{Key Laboratory of Quark and Lepton Physics (MOE) and Institute
of Particle Physics, Central China Normal University, Wuhan 430079, China}

\begin{abstract}
Under the assumption that a quark-gluon plasma droplet is produced and its evolution can be described by hydrodynamics in p + A collisions,  $\gamma$-triggered hadron spectra are studied within a next-to-leading-order perturbative QCD parton model with the medium-modified parton fragmentation functions. The initial conditions and space-time evolution of the small QGP droplet are provided by the superSONIC hydrodynamic model simulations and parton energy loss in such a medium is described by the high-twist (HT) approach. The \textcolor{black}{range of} scaled jet transport coefficient \textcolor{black}{$\hat{q}_0/T_0^3$} in this HT approach is extracted from single hadron suppression in A + A collisions with similar \textcolor{black}{initial medium} temperature as in p + A collisions. Numerical results for this scenario show that $\gamma$-hadron spectra at $p_{\rm T}^\gamma=12-40$ GeV/$c$  are suppressed by 5 $-$ 15\% in the most central 0 - 10\% p + Pb collisions at $\sqrt{s_{\rm NN}}=5.02$ TeV. The suppression becomes weaker at higher transverse momentum of the $\gamma$ trigger. As a comparison, $\gamma$-hadron suppression in Pb + Pb collisions at $\sqrt{s_{\rm NN}}=2.76$ and 5.02 TeV is also predicted.
\end{abstract}

\maketitle

\section{Introduction}

Jet quenching\mbox{\cite{Gyulassy:1990ye,Wang:1991xy,Qin:2015srf}} as reflected in the suppression and azimuthal anisotropy of high $p_{\rm T}$ hadron spectra \cite{Adare:2010sp,Chatrchyan:2012xq,Abelev:2012di,ATLAS:2011ah,Sirunyan:2017pan} are two key evidences for the formation of hot and dense quark-gluon plasma (QGP) in heavy-ion collisions. Recently some phenomena observed in p + Pb collisions seem also to indicate the existence of such small systems of hot and dense medium. For example, the azimuthal anisotropies $v_n$ from two-particle and four-particle correlation measurements in p + Pb collisions at 5.02 TeV \cite{CMS:2012qk,Abelev:2012ola,Aad:2012gla,Aad:2014lta} show a similar behavior of the collective flow as in Pb + Pb collisions. Enhancement of strangeness productions in p + Pb collisions also exhibits similarities to what is observed in Pb + Pb collisions \cite{Abelev:2013haa, Adam:2015vsf}.
However, single charged hadron \cite{ALICE:2012mj,Adam:2014qja,Aad:2015zza,Aad:2016zif,Khachatryan:2015xaa} and single jet spectra \cite{ATLAS:2014cpa,Adam:2015hoa} \textcolor{black}{ practically} do not indicate strong jet quenching phenomena in p + Pb collisions as one would expect if a small droplet of QGP is formed. Similar behavior is also observed for heavy flavor mesons spectra in p + Pb collisions \cite{Sirunyan:2018toe}.

In the experimental study of single hadron and jet suppression, one needs to determine the number of binary collisions for a given class of centrality to calculate the suppression factor relative to the spectra in p + p collisions. This is problematic for p + A collisions due to relatively large dynamical fluctuations of hadron production and leads to large uncertainties \cite{Adam:2014qja}. One can circumvent this problem by measuring the hadron and jet spectra in coincidence with another particle or jet such as the spectra of dihadron, dijet, hadron-jet or $\gamma$-jet. Since one measures the hadron or jet yields per trigger, there is no need to determine the number of binary collisions for normalization. Experimental data  \cite{Chatrchyan:2014hqa} on dijet spectra in p + Pb collisions at  $\sqrt{s_{\rm NN}}=5.02$ TeV, however, show no significant effect of jet quenching within the experimental errors in the nuclear modification of the dijet asymmetry in transverse momentum.
Since trigger biases in dihadron and dijet measurements prefer surface and tangential configurations for coincident production of hadrons or jets \cite{Zhang:2009rn}, the effect of jet quenching should be smaller than in $\gamma$-hadron production where the direct photon does not have strong interaction with the hot medium before being detected. This will be the focus of our study in this paper.

It is generally believed that $\gamma$-jet production is a ``golden probe" for studying parton energy loss since the color-neutral photon does not interact strongly with the hot and dense medium \cite{Adare:2012yt, Afanasiev:2012dg} and can be used to best approximate the transverse momentum of the accompanying jet which is produced together with the photon in the hard processes of the Compton ($qg\rightarrow q\gamma$) or annihilation ($q\bar{q}\rightarrow g\gamma$) scattering \cite{Wang:1996yh, Wang:1996pe}.  Since the produced direct photon does not interact with the hot and dense medium, using it as the coincidence trigger does not lead to biases in the geometrical configuration of the initial production as in the dihadron, hadron-jet or dijet production. Medium modification of $\gamma$-hadron spectra in Au + Au collisions at the Relativistic Heavy-ion Collider (RHIC) \cite{Frantz:2009zn,Adare:2012qi,Abelev:2009gu,STAR:2016jdz} and $\gamma$-jet spectra in Pb + Pb collisions at the Large Hadron Collider (LHC) \cite{Chatrchyan:2012gt,Sirunyan:2017qhf,Sirunyan:2018qec, Aaboud:2018anc} have been observed that are consistent with the picture of jet quenching in the suppression of single hadron and jet spectra. The average fraction of quark jets versus gluon jets in $\gamma$-jet production is larger than that of single and dijet at the same transverse momentum and colliding energy. Quarks lose about half (4/9) less energy than gluons in the QGP medium.
However, phenomenological analyses of experimental data on single jet and $\gamma$-jet suppression show a stronger average jet energy loss for $\gamma$-jets than single inclusive jets \cite{He:2018gks} due to trigger biases. There are less surface and tangential trigger bias in $\gamma$-hadron (jet)  production than in single hadron (jet) and di-hadron (di-jet) production. Therefore, if small droplets of QGP are formed in p + A collisions and energetic partons also experience parton energy loss as in A + A collisions, one should expect to observe more sizable medium modification of $\gamma$-hadron spectra than single hadron and di-hadron spectra in p + A collisions.

To calculate the medium modification of $\gamma$-hadron spectra in p + A collisions, we assume that partons will lose their energy mostly via medium-induced gluon radiation when traversing the medium created in p + A collisions. The radiative jet energy loss is controlled by jet transport coefficient $\hat{q}$ which is also defined as the transverse momentum squared per unit length exchanged between the propagating hard parton and the medium \cite{Baier:1996kr, Baier:1996sk, Baier:1998kq, Guo:2000nz, Wang:2001ifa, Majumder:2009ge}. We will use the \textcolor{black}{range} of the scaled jet transport coefficient \textcolor{black}{$\hat{q}_0/T_0^3$} as extracted from the suppression of single hadron spectra in A + A collisions with similar \textcolor{black}{initial medium} temperature as in p + A collisions.
We will use the initial condition and hydrodynamic evolution of the QGP medium in p + A collisions as provided by the superSONIC hydrodynamic model \cite{Romatschke:2015gxa,Weller:2017tsr}. As comparisons, we also predict the nuclear modification of $\gamma$-hadron spectra in Pb + Pb collisions at the LHC energies.

The remainder of this paper is organized as follows.
In Sec. II, we briefly introduce our framework for the study of the invariant cross section of direct-$\gamma$ and $\gamma$-hadron spectra with large transverse momenta $p_{\rm T}^{\gamma}$ in proton-proton (p + p), proton-nucleus (p + A) and nucleus-nucleus (A + A) collisions. In Sec. III, we numerically calculate the photon spectra in Au + Au collisions at $\sqrt{s_{\rm NN}}=0.2$ TeV, Pb + Pb collisions at $\sqrt{s_{\rm NN}}=2.76$ TeV and 5.02 TeV, and the corresponding photon spectra in p + p collisions as compared with experimental data.
In addition, we will also show the prediction for photon spectra in p + Pb collisions at $\sqrt{s_{\rm NN}}=5.02$ TeV.  In Sec. IV, we focus on cold nuclear matter (CNM) effects on direct photon and $\gamma$-hadron productions without hot medium modification from final state interaction. \textcolor{black}{In Sec. V, we extract the values of jet transport coefficient with different centralities in A + A collisions at both RHIC and LHC energies to estimate uncertainties due to temperature dependence of the scaled jet transport coefficient.} In Sec. VI, we calculate the $\gamma$-triggered fragmentation function $D_{pp}^{\gamma h}(z_{\rm T})$ in p + p collisions and the $\gamma$-hadron suppression factors $I_{AA}^{\gamma h}$ in central Au + Au collisions at 0.2 TeV,  and compare them with experimental data to illustrate the applicability of our model. We also present our predictions for $\gamma$-hadron suppression factors $I_{AA}^{\gamma h}$ in Pb + Pb collisions at $\sqrt{s_{\rm NN}}=$ 2.76 and 5.02 TeV in this section. In Sec. VII, $\gamma$-hadron suppression factors $I_{pA}^{\gamma h}$ for 5.02 TeV p + Pb collisions are shown. A brief summary and discussions are given in Sec. VIII.

\section{pQCD parton model}

\subsection{Direct photon production}
The photon spectrum is the elementary part of the hard processes in high-energy heavy-ion collisions. Photon production is mainly from three processes: (i) quark-gluon Compton scattering $qg\rightarrow q\gamma$, (ii) quark-antiquark annihilation $q\bar{q}\rightarrow g\gamma$, and (iii) photon production from collinear fragmentation of final-state partons. Photons from the first two sources are called ``direct " photons and that from the last source are called ``fragmentation"  photons. The combination of these three sources are called ``prompt" photons \cite{Khachatryan:2010fm, Chatrchyan:2012vq} to differentiate them from photons from hadron decays. The fragmentation photons will be suppressed if an isolation-cut is applied since they are always accompanied by nearly collinear hadrons \cite{Baer:1990ra, Vitev:2008vk}.
For example, such isolation cuts can reduce the fraction of fragmentation photons to less than 10\% for photons with $p _{\rm T}$ smaller than 20 GeV/$c$ in Au + Au collisions at $\sqrt{s_{\rm NN}} =0.2$ TeV \cite{Zhang:2009rn}.
With such isolation cuts it is therefore safe for us to focus mainly on the direct photon production and neglect photons via induced bremsstrahlung.  In addition, we also neglect photons that are produced via jet-photon conversion \cite{Fries:2002kt}. Thermal productions \cite{Srivastava:2008es,Turbide:2005fk} in high-energy heavy-ion collisions are negligible at large transverse momentum as compared to prompt photons.

The differential cross-section of direct photon production in p + p collisions  \cite{Owens:1986mp, Zhou:2010zzm} in perturbative QCD (pQCD) parton model can be expressed as,
\begin{eqnarray}
	\frac{d\sigma_{pp}^{\gamma}}{dy^{\gamma}d^2p_{\rm T}^{\gamma}}&&=\sum_{abd}\int_{x_{a{\rm min}}}^1 dx_a f_{a/p}(x_a,\mu^2) f_{b/p}(x_b,\mu^2)\nonumber \\
&&\times
	\frac{2}{\pi} \frac{x_a x_b}{2x_a-x_{\rm T} e^y}\frac{d\sigma_{ab\rightarrow {\gamma}d}}{d\hat{t}}+\mathcal {O}(\alpha_e \alpha_s^2),
\label{eq:pp-pho}
\end{eqnarray}
where $x_{\rm T}=2p_{\rm T}/\sqrt{s}$, $x_b=x_ax_{\rm T}e^{-y}/(2x_a-x_{\rm T}e^y)$, $x_{a{\rm min}}=x_{\rm T}e^y/(2-x_{\rm T}e^{-y})$, $f_a(x_a,\mu^2)$ is parton distribution functions (PDF's) which we take from CT14 parameterization \cite{Hou:2016nqm} and $d\sigma_{ab\rightarrow \gamma d}/d\hat{t}$ are the tree-level $2 \to 2$ partonic scattering cross sections. The NLO correction at $\mathcal {O}(\alpha_e \alpha_s^2)$ order included in our calculation contains $2\rightarrow2$ virtual diagrams and $2\rightarrow3$ tree diagrams.

Taking into account of the initial-state cold nuclear matter (CNM) effect, one can write down the invariant cross section of direct photon productions in p + A as \cite{Zhou:2010zzm},
\begin{eqnarray}
	\frac{d\sigma_{pA}^{\gamma}}{dy^{\gamma}d^2p_{\rm T}^{\gamma}}&&=\sum_{abd}\int d^2 b\int_{x_{a{\rm min}}}^1 dx_a t_A(\vec{b}) f_{a/A}(x_a,\mu^2,\vec{b})  \nonumber \\
      &&\times
	  f_{b/p}(x_b,\mu^2) \frac{2}{\pi} \frac{x_a x_b}{2x_a-x_{\rm T} e^y} \frac{d\sigma_{ab\rightarrow {\gamma}d}}{d\hat{t}} \nonumber \\
	  &&
	  +\mathcal {O}(\alpha_e \alpha_s^2) ,
\label{eq:pA-pho}
\end{eqnarray}
where $ t_A(\vec{b})$ is the nuclear thickness function at an impact-parameter $\vec b$ given by the Woods-Saxon distribution \cite{Jacobs:2000wy}. Since one of the incoming partons comes from a nucleus, the PDF in the nuclear target should be the nuclear modified PDF $f_{a/A}(x_a,\mu^2,\textcolor{black}{\vec{b}})$ \cite{Wang:1996yf,Li:2001xa}:
\begin{eqnarray}
f_{a/A}(x_a,\mu^2,\vec{b}) &&= S_{a/A}(x_a,\mu^2,\vec{b})\left[\frac{Z}{A}f_{a/p}(x_a,\mu^2)\right. \nonumber\\
&&+\left.\left(1-\frac{Z}{A}\right)f_{a/n}(x_a,\mu^2)\right],
\end{eqnarray}
where $Z$ and $A$ are the charge and mass number of the nucleus, respectively.
 The nuclear modification factor$S_{a/A}(x_a,\mu^2,\textcolor{black}{\vec{b}})$ of the PDFs will be given by the EPPS16 \cite{Eskola:2016oht} parameterization.

In A + A collisions, the  invariant cross section of direct photon production at high transverse momentum may be obtained as \cite{Zhou:2010zzm},
\begin{eqnarray}
	\frac{d\sigma_{A \textcolor{black}{B}}^{\gamma}}{dy^{\gamma}d^2p_{\rm T}^{\gamma}}&&=\sum_{abd}\int d^2b \int d^2 r\int_{x_{a{\rm min}}}^1 dx_a t_A(\vec{r}) t_B(\vec{r}+\vec{b}) \nonumber \\
	&&\times f_{a/A}(x_a,\mu^2,\vec{r})  f_{b/B}(x_a,\mu^2,\vec{r}+\vec{b})\nonumber \\
     &&\times \frac{2}{\pi} \frac{x_a x_b}{2x_a-x_{\rm T} e^y} \frac{d\sigma_{ab\rightarrow {\gamma}d}}{d\hat{t}} +\mathcal {O}(\alpha_e \alpha_s^2),
\label{eq:AA-pho}
\end{eqnarray}
where the range of the integration over the impact-parameter $b$ is specified by the range of centralities in A + A collisions.

Using the spectrum in p + p collisions as a baseline, the nuclear modification factor of direct photon productions in p + A collisions can be defined as \cite{Wang:1991xy,Wang:1998ww,Wang:2004yv},
\begin{eqnarray}
R_{pA}^{\gamma}=\frac{d\sigma_{pA}^{\gamma}/dy^{\gamma}d^2p_{\rm T}^{\gamma}}{\langle N^{pA}_{\rm binary}\rangle d{\sigma}_{pp}^{\gamma}/dy^{\gamma}d^2p_{\rm T}^{\gamma}}.
\label{eq:R_pA}
\end{eqnarray}
where $\langle N^{pA}_{\rm binary}\rangle =\int d^2b t_A(\vec{b})$ for p + A collisions. In A + A collisions, the nuclear modification factor for direct photon production is similarly defined as \cite{Wang:1991xy,Wang:1998ww,Wang:2004yv},
\begin{eqnarray}
R_{A \textcolor{black}{B}}^{\gamma}=\frac{d\sigma_{A\textcolor{black}{B}}^{\gamma}/dy^{\gamma}d^2p_{\rm T}^{\gamma}}{\langle  N^{A \textcolor{black}{B}}_{\rm binary}  \rangle d{\sigma}_{pp}^{\gamma}/dy^{\gamma}d^2p_{\rm T}^{\gamma}}.
\label{eq:R_AA}
\end{eqnarray}
where $\langle  N^{A\textcolor{black}{B}}_{\rm binary}  \rangle =\int d^2b T_{A\textcolor{black}{B}}(\vec b)$ and $T_{A\textcolor{black}{B}}(\vec{b})=\int d^2 r  t_A(\vec{r}) t_B(\vec{r}+\vec{b})$ is the overlap function of two colliding nuclei. Since direct photons do not have final state interaction, we only need to take into account CNM effect on the initial parton distributions.

In the calculation of direct photon and photon-hadron spectra in both p + A and A + A collisions, the range of the integration over the impact-parameter is specified by the range of event centralities of the collisions. The centralities in our calculations are classified according to the percentile  of impact-parameter distribution of the total cross section which are matched to the centralities in experimental data defined by the percentile event distribution in charged hadron rapidity density $dN_{ch}/dy$. We will use event-by-event hydrodynamic simulations of the space-time evolution of the bulk medium for the calculation of parton energy loss and the modified jet fragmentation functions. For each centrality bin, the hydro we used is averaged over 200 events.

\subsection{$\gamma$-hadron spectra}

If the contributions from fragmentation photons are neglected, the invariant cross section of $\gamma$-hadron production only involves the fragmentation function of one parton to a hadron. In p + p collisions, the cross section of  $\gamma$-hadron can be expressed as \cite{Owens:1986mp},
\begin{eqnarray}
	\frac{d\sigma_{pp}^{\gamma h}}{dy^{\gamma}d^2p_{\rm T}^{\gamma} dy^{h} d^2p_{\rm T}^{h}}&&=\sum_{abd}\int dz_d f_{a/p}(x_a,\mu^2) f_{b/p}(x_b,\mu^2) \nonumber \\
	&& \times \frac{x_ax_b}{\pi z_d^2} \frac{d\sigma_{ab\rightarrow {\gamma}d}}{d\hat{t}} D_{h/d}(z_d,\mu^2) \nonumber \\
	&& \times \delta^2(\vec{p}_{\rm T}^{~\gamma}+\frac{\vec{p}_{\rm T}^{~h}}{z_d})\ + {O}(\alpha_e \alpha_s^2),
\label{eq:pp-pho-h}
\end{eqnarray}
where $z_d=p_{{\rm T}h}/p_{{\rm T}d}$. We use the Kniehl-Kramer-Potter parametrization \cite{Kniehl:2000fe} for the vacuum fragmentation function $D_{h/d}(z_d,\mu^2)$.

Similarly, the invariant cross section of $\gamma$-hadron productions in p + A collisions can be written as,
\begin{eqnarray}
	\frac{d\sigma_{pA}^{\gamma h}}{dy^{\gamma}d^2p_{\rm T}^{\gamma} dy^{h} d^2p_{\rm T}^{h}}&&=\sum_{abd}\int  d^2 b \frac{d\phi_b}{2\pi} dz_d t_A(\vec{b})  f_{b/p}(x_b,\mu^2) \nonumber \\
	&& \times f_{a/A}(x_a,\mu^2,\vec{b}) \frac{x_a x_b}{\pi z_d^2} \frac{d\sigma_{ab\rightarrow {\gamma}d}}{d\hat{t}} \nonumber \\
	&& \times \tilde{D}_{h/d}(z_d,\mu^2,\Delta{E_d}) \nonumber\\
	&& \times \delta^2(\vec{p}_{\rm T}^{~\gamma}+\frac{\vec{p}_{\rm T}^{~h}}{z_d})\ + {O}(\alpha_e \alpha_s^2),
\label{eq:pA-pho-h}
\end{eqnarray}
while the cross section of $\gamma$-hadron productions in A + A collisions can be expressed as,
\begin{eqnarray}
	\frac{d\sigma_{A \textcolor{black}{B}}^{\gamma h}}{dy^{\gamma}d^2p_{\rm T}^{\gamma} dy^{h} d^2p_{\rm T}^{h}}&&=\sum_{abd}\int d^2b d^2 r \int \frac{d\phi_b}{2\pi} dz_d t_A(\vec{r})\nonumber \\
	 && \times t_B(\vec{r}+\vec{b}) f_{a/A}(x_a,\mu^2,\vec{r})\nonumber \\
	 && \times f_{b/B}(x_b,\mu^2,\vec{r}+\vec{b})
	 \frac{x_a x_b}{\pi z_d^2} \frac{d\sigma_{ab\rightarrow {\gamma}d}}{d\hat{t}} \nonumber \\
	&& \times \tilde{D}_{h/d}(z_d,\mu^2,\Delta{E_d}) \nonumber\\
	&& \times \delta^2(\vec{p}_{\rm T}^{~\gamma}+\frac{\vec{p}_{\rm T}^{~h}}{z_d})\ + {O}(\alpha_e \alpha_s^2),
\label{eq:AA-pho-h}
\end{eqnarray}
where $\phi_b$ is the azimuthal angle between the parton's propagating direction $\vec n$ and the impact-parameter $\vec b$.
The medium-modified fragmentation function $\tilde{D}_{h/d}(z_d,\mu^2,\Delta{E_d})$ can be calculated as \cite{Wang:2004yv,Zhang:2007ja,Zhang:2009rn},
\begin{eqnarray}
	&& \tilde{D}_{h/d}(z_d,\mu^2,\Delta{E_d}) = (1-e^{-\langle{N_g^d}\rangle})\left[\frac{z'_d}{z_d}D_{h/d}(z'_d,\mu^2)\right. \nonumber\\
&& ~~+\left.{\langle{N_g^d}\rangle}\frac{{z_g}'}{z_d}D_{h/g}({z_g}',\mu^2)\right]+e^{-\langle{N_g^d}\rangle}D_{h/d}({z_d},\mu^2),
\label{eq:mFF}
\end{eqnarray}
where ${z_d}'=p_{\rm T}/(p_{{\rm T}d}-\Delta{E_d})$ is the momentum fraction of a hadron with transverse momentum $p_{\rm T}$ from a parton with initial transverse momentum  $p_{{\rm T}d}$ that has lost  energy $\Delta{E}_d$ while propagating through the hot medium, $z_d=p_{\rm T}/p_{{\rm T}d}$ is the hadron's momentum fraction when the parton fragments in the vacuum. $\langle{N_g^d}\rangle$ is the averaged number of radiated gluons and ${z_g}'=\langle{N_g^d}\rangle p_{\rm T}/\Delta{E_d}$ is the momentum fraction of a hadron from a radiated gluon who carries an average energy $\Delta{E}_d/\langle{N_g^d}\rangle$. The factor $e^{-\langle{N_g^d}\rangle}$ is the probability for a parton to propagate through the medium without suffering any inelastic scattering. Correspondingly, $(1-e^{-\langle{N_g^d}\rangle})$ is the probability for a parton to suffer at least one inelastic scattering.

Within the high-twist formalism \cite{Wang:2001ifa, Wang:2009qb, Wang:2001cs, Wang:2002ri}, the radiative energy loss for a parton $d$ with initial energy $E$ can be calculated as an integral over the propagation path,
\begin{eqnarray}
\frac{\Delta{E_d}}{E} &&= \frac{2C_A\alpha_s}{\pi} \int d\tau \int \frac{dl_{\rm T}^2}{l_{\rm T}^4}\int dz  \left[1+(1-z)^2\right]  \nonumber\\
	&&\times \hat{q}_d(\tau, \vec r+(\tau-\tau_0)\vec n) \sin^2\left[\frac{l_{\rm T}^2(\tau-\tau_0)}{4z(1-z)E}\right],
\label{eq:deltaE}
\end{eqnarray}
starting at an initial time $\tau_0$, where $C_A=3$, and $\alpha_s$ is the strong coupling constant, $l_{\rm T}$ is the transverse momentum of radiated gluon and  $z$ \textcolor{black}{is its longitudinal momentum fraction.}  Note that the jet transport coefficient for a gluon and a quark is related by a constant color factor $\hat q_A/\hat q_F=C_A/C_F$. Therefore the energy loss of a gluon is simply $C_A/C_F$ times that of a quark \cite{Wang:2009qb}.
The average number of radiated gluons from the propagating hard parton $d$ is \cite{Chang:2014fba},
\begin{eqnarray}
	\langle N_g^d \rangle &&= \frac{2C_A \alpha_{s}}{\pi}\int d\tau \int \frac{dl_{\rm T}^2}{l_{\rm T}^4}\int \frac{dz}{z}  \left[1+(1-z)^2\right]  \nonumber\\
	&&\times \hat{q}_d(\tau, \vec r+(\tau-\tau_0)\vec n) \sin^2(\frac{l_{\rm T}^2(\tau-\tau_0)}{4z(1-z)E}).
\label{eq:deltaNg}
\end{eqnarray}

We also assume the jet transport parameter have the following temperature scaling and dependence on the fluid velocity \citep{Baier:1996sk}\cite{Chen:2010te},
\begin{eqnarray}
\hat q_d = \hat q_{d0} \frac{T^3}{T_0^3}\frac{p^{\mu}\cdot u_{\mu}}{p_0},
\label{eq:qhat}
\end{eqnarray}
where $p^{\mu}=(p_0,\vec p)$ is the four momentum of the parton, $u^{\mu}$ is the local four flow velocity of the fluid, $T$ is the local temperature of the medium and $T_0$ is a reference temperature which is usually taken as the highest temperature at the center of the medium at the initial time $\tau_0$ for each collision centrality in proton-nucleus or nucleus-nucleus collisions. In our study here we will vary  $\tau_0$ to explore the sensitivity of parton energy loss on the initial time in p + Pb collisions.

We assume that QGP is also formed in high energy p + A collisions and its evolution and final bulk hadron production can be governed by the same hydrodynamics as in A + A collisions. \textcolor{black}{We also assume the same scaled jet transport coefficient $\hat{q}_0/T_0^3$ in p + Pb collisions at $5.02$ TeV as extracted from single inclusive hadron spectra in Au + Au collisions at 0.2 TeV and in peripheral Pb + Pb collisions at 2.76 and 5.02 TeV in the range of the initial temperature at the center of the QGP droplet as given by the hydrodynamic model.}

From $\gamma$-hadron spectra at high transverse momentum in p + p collisions, we define the $\gamma$-triggered fragmentation function (FF)  $D_{pp}^{\gamma h}(z_{\rm T})$  as \cite{Wang:2003aw},
\begin{eqnarray}
D_{pp}^{\gamma h}(z_{\rm T})=p_{\rm T}^{\gamma}\frac{d{\sigma}_{pp}^{\gamma h}/dy^{\gamma}dp_{\rm T}^{\gamma}dy^{h}dp_{\rm T}^{h}}{d{\sigma}_{pp}^{\gamma}/dy^{\gamma}dp_{\rm T}^{\gamma}}.
\label{eq:D_pp}
\end{eqnarray}
In p + A collisions, $D_{pA}^{\gamma h}(z_{\rm T})$  is defined as \cite{Wang:2003aw},
\begin{eqnarray}
D_{pA}^{\gamma h}(z_{\rm T})=p_{\rm T}^{\gamma}\frac{d{\sigma}_{pA}^{\gamma h}/dy^{\gamma}dp_{\rm T}^{\gamma}dy^{h}dp_{\rm T}^{h}}{d{\sigma}_{pA}^{\gamma}/dy^{\gamma}dp_{\rm T}^{\gamma}},
\label{eq:D_pA}
\end{eqnarray}
where the numerator is $\gamma$-hadron cross section and the denominator is the cross section of photon production.  Similarly in A + A collisions, $\gamma$-triggered fragmentation function is defined as,
\begin{eqnarray}
	D_{A\textcolor{black}{B}}^{\gamma h}(z_{\rm T})=p_{\rm T}^{\gamma}\frac{d\sigma_{A\textcolor{black}{B}}^{\gamma h}/dy^{\gamma}dp_{\rm T}^{\gamma}dy^{h}dp_{\rm T}^{h}}{d{\sigma}_{A \textcolor{black}{B}}^{\gamma}/dy^{\gamma}dp_{\rm T}^{\gamma}}. \ \ \ \
\label{eq:D_AA}
\end{eqnarray}
In the numerical calculations to be compared to experimental data, one has to integrate the kinematics over the experimental coverage including the opening angle between the hadron and photon. The nuclear modification factor of the triggered fragmentation function $I_{pA}^{\gamma h}$ as a function of $z_{\rm T}=p_{\rm T}^{h}/p_{\rm T}^{\gamma}$ can be defined as \cite{Zhang:2007ja},
\begin{eqnarray}
 I_{pA}^{\gamma h}(z_{\rm T}) = \frac{D_{pA}^{\gamma h}(z_{\rm T})}{D_{pp}^{\gamma h}(z_{\rm T})},
 \label{eq:I_pA}
\end{eqnarray}
which can be similarly defined for A + A collisions.

Furthermore, $I_{p A}^{\gamma h}(z_{\rm T})$ can be rewritten in the following form,
\begin{eqnarray}
 I_{pA}^{\gamma h}(z_{\rm T}) = \frac{J_{pA}^{\gamma h}(z_{\rm T})}{R_{p A}^{\gamma}(p_{\rm T})},
\label{eq:IpA-JpA}
\end{eqnarray}
where $J_{pA}^{\gamma h}$ is the ratio of $\gamma$-hadron yield in p + A collisions over that in p + p collisions,
\begin{eqnarray}
 J_{pA}^{\gamma h}(z_{\rm T}) =\frac{\frac {d{\sigma}_{pA}^{\gamma h}}{dy^{\gamma}dp_{\rm T}^{\gamma}dy^{h}dp_{\rm T}^{h}d \phi}}{\langle N^{pA}_{\rm binary}\rangle \frac {d{\sigma}_{pp}^{\gamma h}}{dy^{\gamma}dp_{\rm T}^{\gamma}dy^{h}dp_{\rm T}^{h}d \phi}},
\label{eq:J_pA}
\end{eqnarray}
without normalization by the production cross section of the trigger photon.  \textcolor{black}{Eq. (\ref{eq:IpA-JpA}) is just a different form of Eq.~(\ref{eq:I_pA}), which is expressed in terms of two modification factors. This way we can isolate the cold nuclear modification factor of the trigger photons from the double differential cross section for the trigger photon and hadron and see its effect in $I^{\gamma h}_{pA}$.}  In the absence of any CNM effect on direct photon spectra, i.e.,  $R_{pA}^{\gamma}(p_{\rm T})=1$, then $I_{pA}^{\gamma h}(z_{\rm T}) = J_{pA}^{\gamma h}(z_{\rm T})$.

\textcolor{black}{Correspondingly}, in A + A collisions the ratio of $\gamma$-hadron yield over that in p + p collisions can be written as,
\begin{eqnarray}
 J_{A\textcolor{black}{B}}^{\gamma h}(z_{\rm T}) =\frac{\frac {d\sigma_{A\textcolor{black}{B}}^{\gamma h}}{dy^{\gamma}dp_{\rm T}^{\gamma}dy^{h}dp_{\rm T}^{h}d \phi}}{\langle N^{A\textcolor{black}{B}}_{\rm binary}\rangle \frac {d{\sigma}_{pp}^{\gamma h}}{dy^{\gamma}dp_{\rm T}^{\gamma}dy^{h}dp_{\rm T}^{h}d \phi}}.
\label{eq:J_AA}
\end{eqnarray}

\section{Direct photon production cross section}

\begin{figure}[tbh!]
\vspace{-5mm}
\begin{center}
\includegraphics[width=0.4\textwidth]{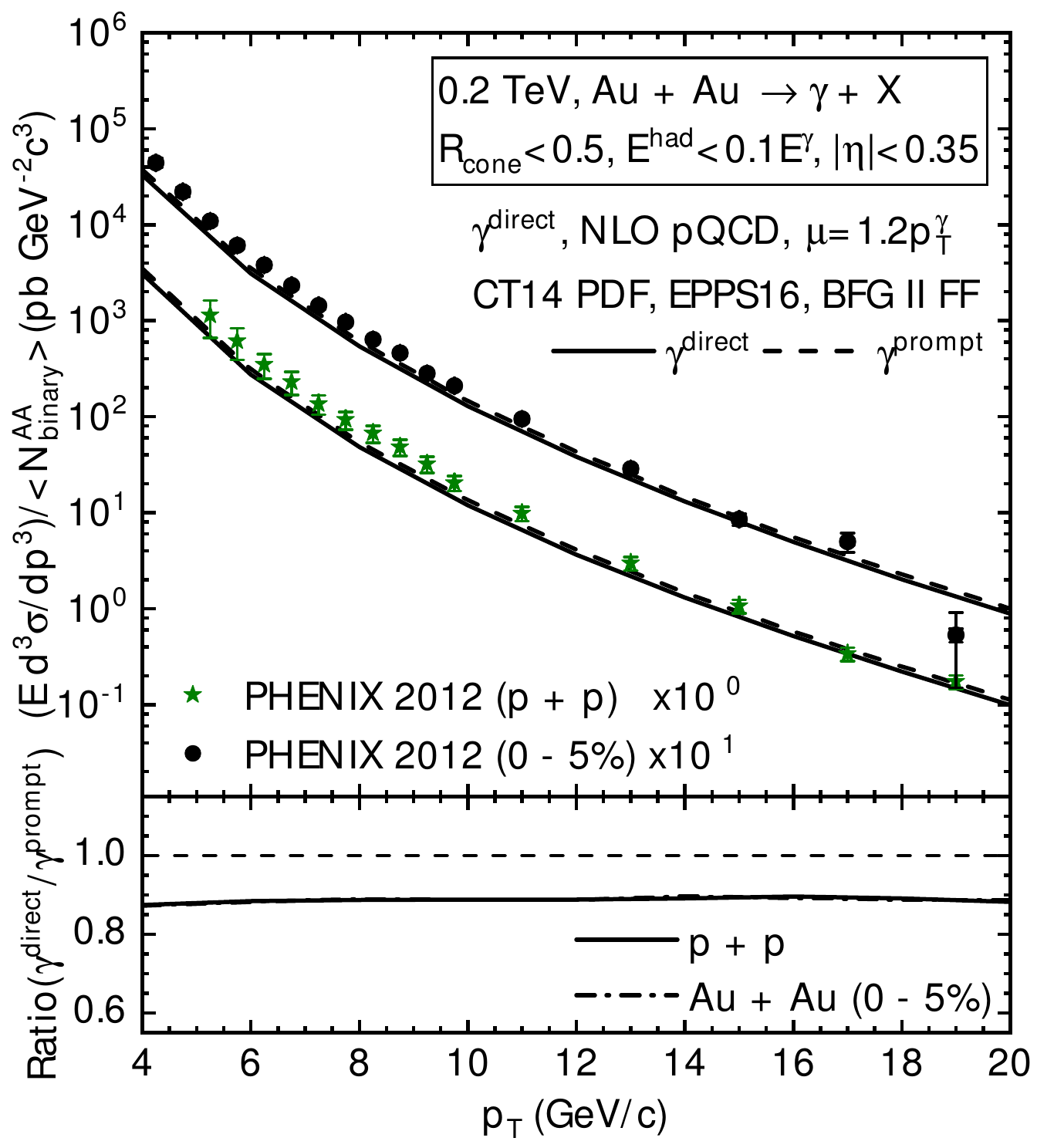}
\end{center}
\vspace{-5mm}
	\caption[*]{Direct photon (solid) and prompt photon (dashed) spectra as a function of $p_{\rm T}^{\gamma}$ for 0 - 5\% Au + Au collisions (scaled by $\langle N_{\rm binary}^{AA}\rangle$) and p + p collisions at $\sqrt{s_{\rm NN}}=0.2$ TeV, scaled by the factors for easier viewing, as compared with PHENIX data \cite{Adare:2012yt,Afanasiev:2012dg}. The ratio of contributions of direct photon to prompt photon productions for 0 - 5\% Au + Au collisions and p + p collisions are shown in lower panel.}
\label{fig:pho-AuAu-200}
\end{figure}
\begin{figure}[tbh!]
\begin{center}
\includegraphics[width=0.4\textwidth]{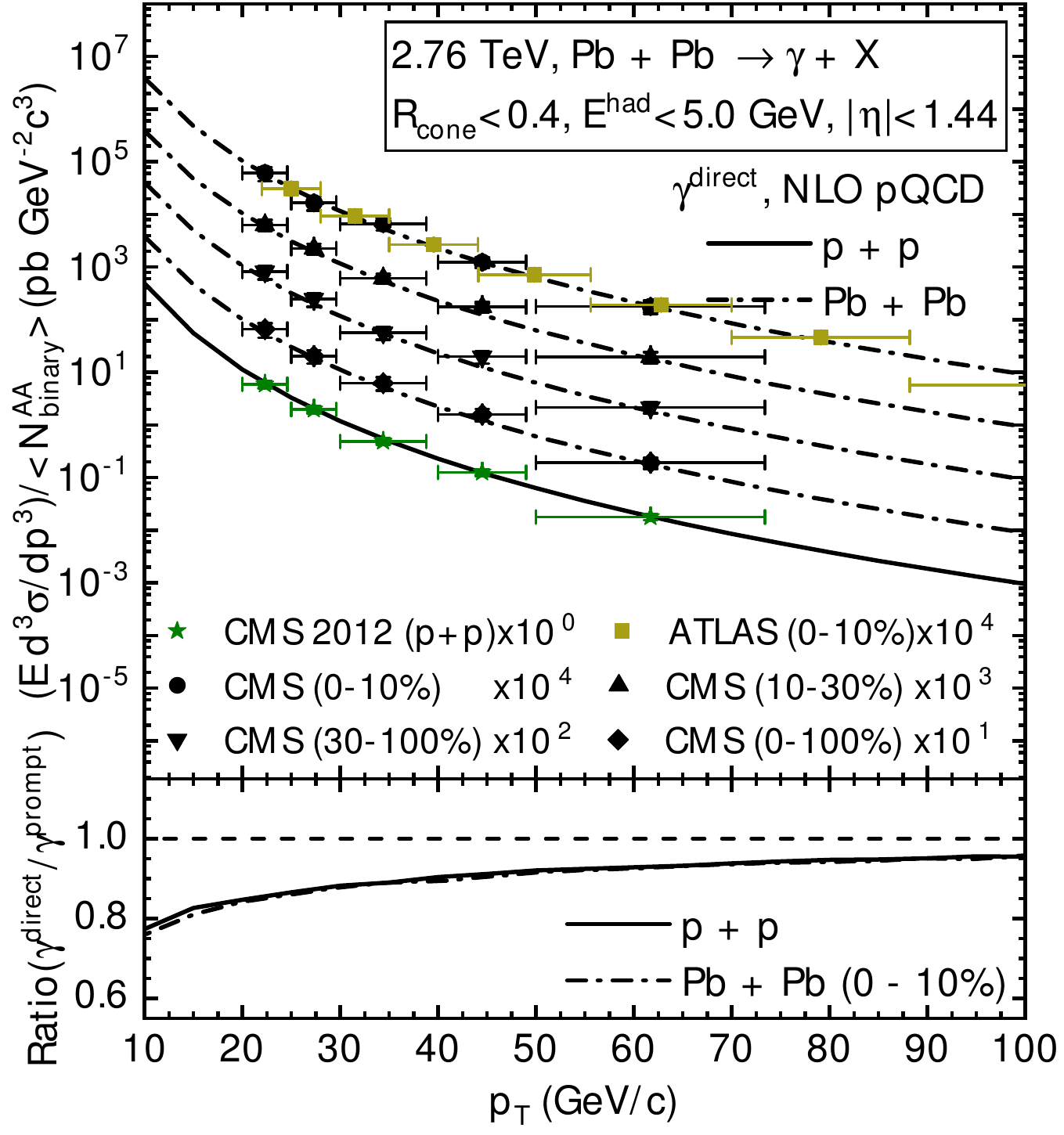}
\end{center}
\vspace{-5mm}
	\caption[*]{Direct photon spectra as a function of $p_{\rm T}^{\gamma}$ for 0 - 10\%, 10 - 30\%, 30 - 100\%, 0 - 100\% Pb + Pb collisions (dot-dashed) (scaled by $\langle N_{\rm binary}^{AA}\rangle$) and p + p collisions (solid) at $\sqrt{s_{\rm NN}}=$ 2.76 TeV, scaled by the factors for easier viewing, as compared with experimental data \cite{Chatrchyan:2012vq,Aad:2015lcb}. The ratio of contributions of direct photon to prompt photon productions for 0 - 10\% Pb + Pb collisions and p + p collisions are shown in lower panel.}
\label{fig:pho-pbpb-2760}
\end{figure}
\begin{figure}
\vspace{-5mm}
\begin{center}
\includegraphics[width=0.4\textwidth]{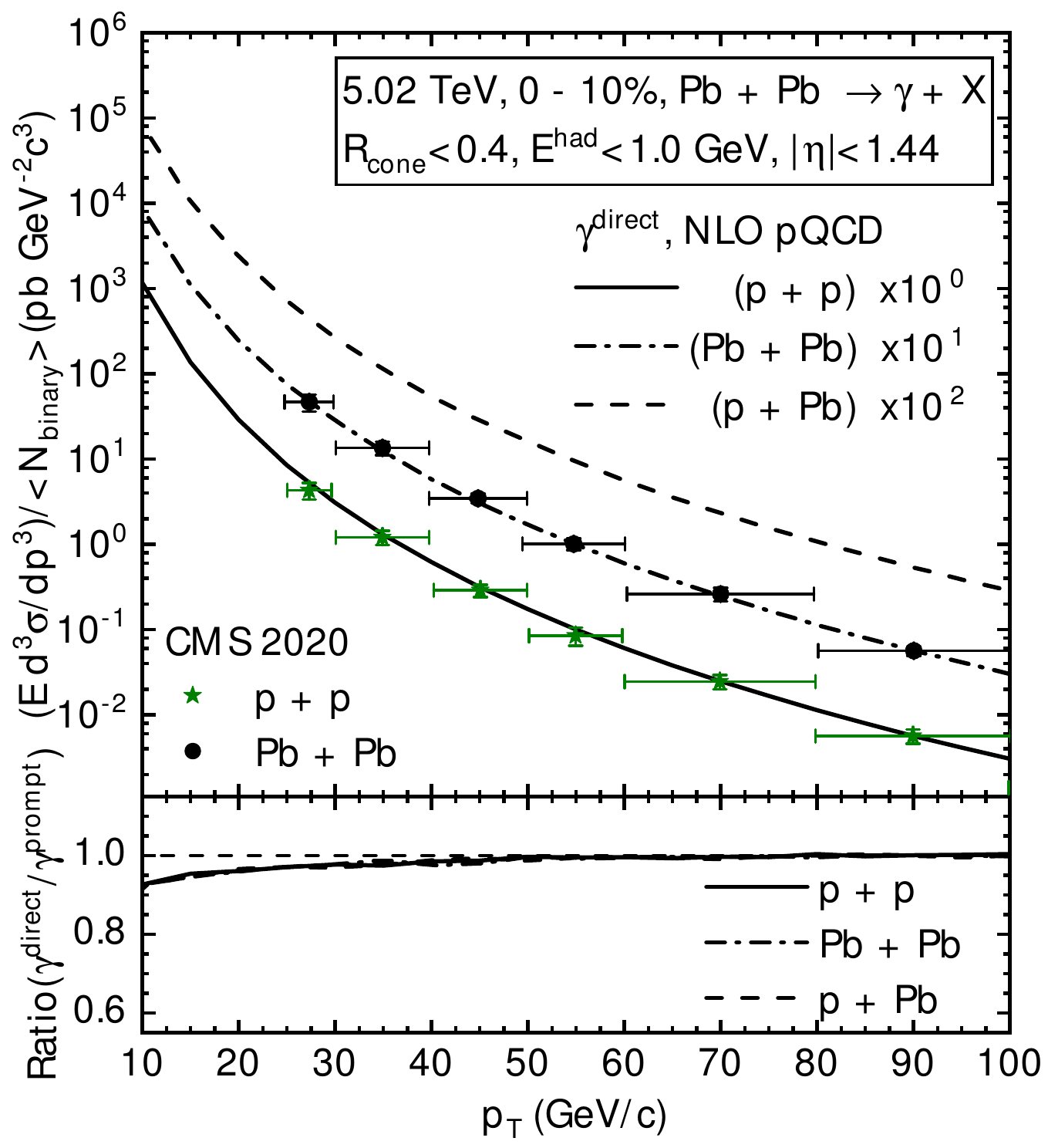}
\end{center}
\vspace{-5mm}
	\caption[*]{Direct photon spectra as a function of $p_{\rm T}^{\gamma}$ for 0 - 10\% Pb + Pb collisions (dot-dashed) (scaled by $\langle N_{\rm binary}^{AA}\rangle$) and p + p collisions (solid) at $\sqrt{s_{\rm NN}}=5.02$ TeV as compared with CMS data \cite{Sirunyan:2020ycu}, and the prediction for direct photon spectrum for 0 - 10\% p + Pb collisions (dashed) (scaled by $\langle N_{\rm binary}^{pA}\rangle$) at $\sqrt{s_{\rm NN}}=5.02$ TeV, scaled by different factors for easier viewing. The ratio of contributions of direct photon to prompt photon productions for these three collisions are shown in lower panel.}
\label{fig:pho-ppb-pbpb}
\end{figure}

The numerical results for the photon invariant cross section in central Au + Au collisions (scaled by $\langle N_{binary}^{AA} \rangle$) and p + p collisions at $\sqrt{s_{\rm NN}}=0.2$ TeV are compared with PHENIX data \cite{Adare:2012yt, Afanasiev:2012dg} in Fig. \ref{fig:pho-AuAu-200}. The cross sections of direct photon and prompt photon are both shown in this figure and their ratios are shown in the lower panel. \textcolor{black}{For fragmentation functions of prompt photons, we use the BFG II FFs \cite{Bourhis:1997yu} there.} The pQCD parton model can describe the experiment data well. With isolation cuts ($R_{\rm cone}<0.5$, $E^{had}<0.1E^{\gamma}$) contributions of the fragmentation photons are about 10\% both in p + p and 0 - 5\% Au + Au collisions at 0.2 TeV.

We also show the direct photon spectra in 0 - 10\%, 10 - 30\%, 30 - 100\%, 0 - 100\% Pb + Pb collisions (scaled by $\langle N_{binary}^{AA} \rangle$) and p + p collisions at $\sqrt{s_{\rm NN}}=2.76$ TeV as compared with experimental data from CMS and ATLAS \cite{Chatrchyan:2012vq,Aad:2015lcb} in Fig. \ref{fig:pho-pbpb-2760}. The pQCD parton model results are in good agreement with the experimental data. In the lower panel of Fig. \ref{fig:pho-pbpb-2760}, the ratios of direct photons to prompt photons with isolation cuts ($R_{\rm cone}<0.4$, $E^{had}<5.0$~GeV) for p + p collisions and 0 - 10\%  Pb + Pb collisions are shown to vary from about 80\% - 90\%.  The contributions of fragmentation photons become smaller at larger $p_{\rm T}^{\gamma}$ and it is less than 10\% for $p_{\rm T}^{\gamma} > 50$ GeV/$c$.

Finally in Fig. \ref{fig:pho-ppb-pbpb}, the direct photon spectra from pQCD model as a function of $p_{\rm T}^{\gamma}$ in 0 - 10\% Pb + Pb collisions (scaled by $\langle N_{binary}^{AA} \rangle$) and p + p collisions at $\sqrt{s_{\rm NN}}=5.02$ TeV are compared with CMS data \cite{Sirunyan:2020ycu}. The prediction for direct photon spectrum (scaled by $\langle N_{binary}^{pA} \rangle$) for 0 - 10\% p + Pb collisions at 5.02 TeV are also shown.  With isolation cuts ($R_{\rm cone}<0.4$, $E^{had}<1.0$~GeV) the contributions of direct photons to prompt photons for p + p collisions, 0 - 10\%  Pb + Pb collisions and 0 - 10\%  p + Pb collisions are also shown in the lower panel. Compared to Fig. \ref{fig:pho-pbpb-2760}, the contributions of fragmentation photons are greatly reduced as the selection (isolation cuts) conditions become more strict, and it becomes negligible for $p_{\rm T}^{\gamma} > 20$ GeV/$c$.
One can, therefore, neglect the contributions of fragmentation photons in numerical calculations with such isolation cuts in the following.

\section{$\gamma$-hadron spectra and CNM effects}

To study the net suppressions of $\gamma$-hadron spectra caused by jet quenching, we need to examine the cold nuclear matter (CNM) effect on $\gamma$-hadron spectra first. We study in this section, the CNM effects on both $\gamma$ spectra and $\gamma$-hadron spectra in A + A and p + A collisions without the effect of the hot QGP medium. To turn off the effect of hot QGP medium in $\gamma$-hadron spectra, we simply replace the medium modified fragmentation function $\tilde{D}_{h/d}(z_d,\mu^2,\Delta{E_d})$ with the vacuum one $D_{h/d}(z_d,\mu^2)$ in p + A [Eq. (\ref{eq:pA-pho-h})]
and  A  + A [Eq. (\ref{eq:AA-pho-h})] collisions.

\begin{figure}[tbh]
\begin{center}
\includegraphics[width=0.4\textwidth]{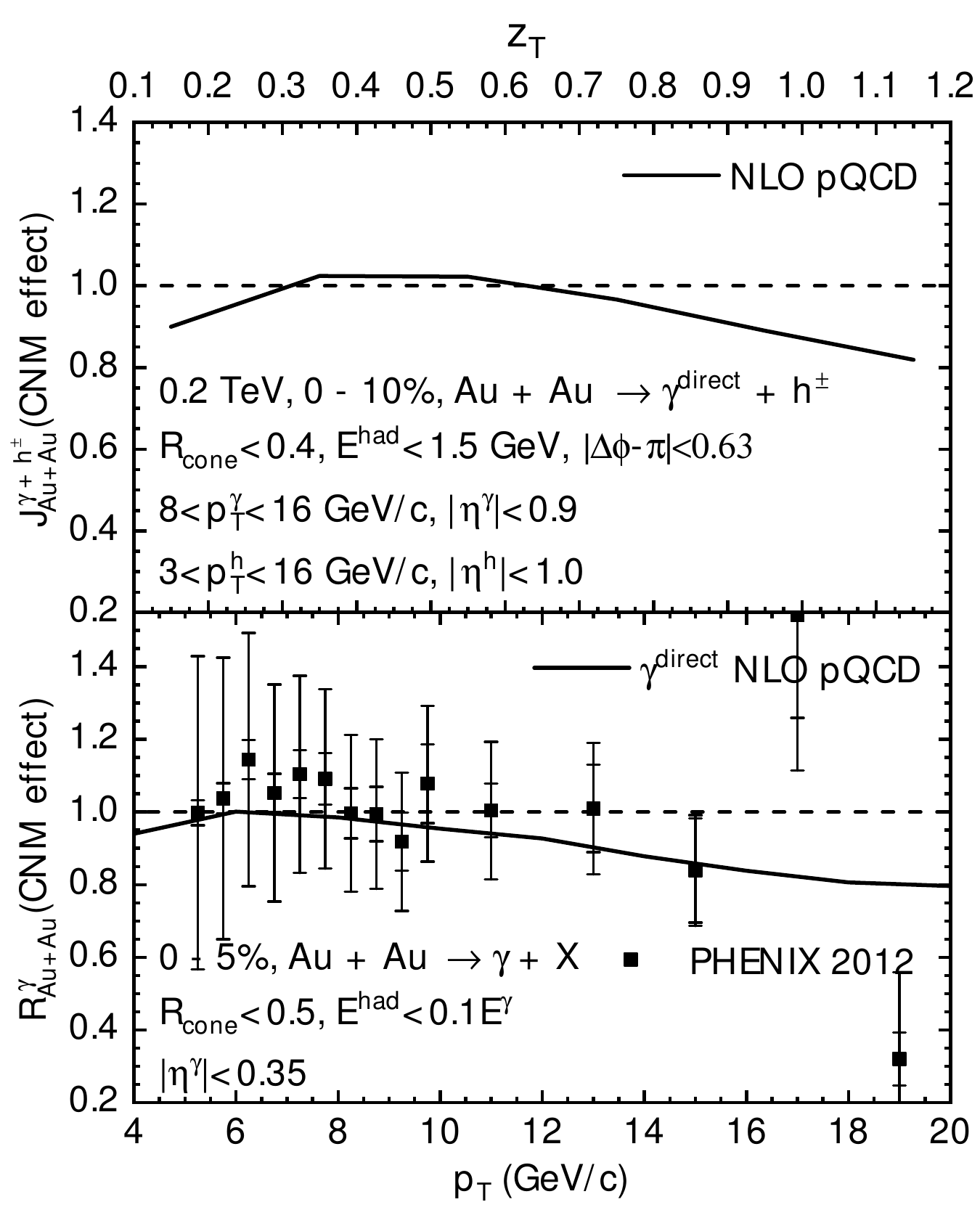}
\end{center}
\vspace{-5mm}
	\caption[*]{The modification factor due to cold nuclear matter (CNM) effect on $\gamma^{\rm dir}$-hadron spectra with ($8<p_{\rm T}^{\gamma}<16$ GeV/$c$,  $3<p_{\rm T}^{h}<16$ GeV/$c$) in 0 - 10\% Au + Au collisions  and on direct photon productions in 0 - 5\% Au + Au collisions as compared with PHENIX data \cite{Afanasiev:2012dg} at $\sqrt{s_{\rm NN}}=0.2$ TeV. The  $\gamma^{\rm dir}$-hadron suppression factor $J_{AA}^{\gamma h}(z_{\rm T})$ without normalization by the number of trigger photons is shown in the upper panel while the direct photon suppression factor $R_{AA}^{\gamma}(p_{\rm T}^{\gamma})$ is shown in the lower panel.}
\label{fig:CNM-AuAu}
\end{figure}

In the upper panel of Fig. \ref{fig:CNM-AuAu}, we show our calculations of $\gamma^{\rm dir}$-hadron cold nuclear modification factor $J_{AA}^{\gamma h}$(without normalization by the number of trigger photons ) as a function of $z_{\rm T}=p_{\rm T}^h/p_{\rm T}^{\gamma}$ (with $8<p_{\rm T}^{\gamma}<16$ GeV/$c$) in 0 - 10\% Au + Au collisions at $\sqrt{s_{\rm NN}}=$ 0.2 TeV.
In the lower panel is direct photon modification factor $R_{AA}^{\gamma}$ by cold nuclear matter  for 0 - 5\% Au + Au collisions at 0.2 TeV which agrees with the experimental data \cite{Afanasiev:2012dg} well.
One can see that when photon $p_{\rm T}$ is less than 10 GeV/$c$, there is not significant cold nuclear matter effect. However, for $p_{\rm T}^{\gamma} >10$ GeV/$c$ or $z_{\rm T}>0.7$, both the photon and $\gamma$-hadron spectra are suppressed due to the EMC effect in the nuclear parton distribution functions \cite{Eskola:2016oht,Eskola:2009uj}.
Taking the results in these two panels together, we find the CNM effect on $\gamma$-hadron spectra suppression $I_{AA}^{\gamma h}$, which is normalized by the number of trigger photons, will have a slight enhancement according to the A + A counterpart of Eq.~(\ref{eq:IpA-JpA}).

Similarly,  $\gamma^{\rm dir}$-hadron and direct photon spectra in 0 - 10\% Pb + Pb collisions at $\sqrt{s_{\rm NN}}=2.76$ TeV and in 0 - 10\% p + Pb collisions at $\sqrt{s_{\rm NN}}=5.02$ TeV are shown in the left and right panels, respectively,  of Fig. \ref{fig:CNM-pb-ppb}. We have also calculated the nuclear modification of $\gamma$-hadron spectra due to CNM effect in Pb + Pb collisions at 5.02 TeV. The result is very similar to that in Pb + Pb collisions at $\sqrt{s_{\rm NN}}=2.76$ TeV as shown in Fig. \ref{fig:CNM-pb-ppb}.
The $\gamma^{\rm dir}$-hadron cold nuclear modification factors $J_{PbPb}^{\gamma h}$  and $J_{pPb}^{\gamma h}$ with $12<p_{\rm T}^{\gamma}<40$ GeV/$c$ (without normalization by the trigger photon yields) as a function of $z_{\rm T}$ \textcolor{black}{are} approximately equal to one as shown in the  upper panels of Fig. \ref{fig:CNM-pb-ppb}. The direct photon modification factor $R^{\gamma}$ due to cold nuclear effect is smaller than one for $p_{\rm T}^{\gamma}<35$ GeV/$c$ both in Pb + Pb and p + Pb collisions, where the Bjorken $x$ of the initial-state parton falls in the region of nuclear shadowing \cite{Eskola:2016oht,Eskola:2009uj}.
At an average value of photon trigger transverse momentum $p_{\rm T}^{\gamma}=26$ GeV/$c$, the direct photon spectrum has a suppression of about 10\% which causes the $\gamma$-hadron modification factor $I^{\gamma h}$ becoming a little larger than one. \textcolor{black}{As supplements to Fig. \ref{fig:CNM-AuAu} and \ref{fig:CNM-pb-ppb}, we show the corresponding $I_{AA}^{\gamma h}$ and $I_{pPb}^{\gamma h}$ with only CNM effect in Appendix A.}
One can conclude that the medium modification factors $I_{PbPb}^{\gamma h}$ and $I_{pPb}^{\gamma h}$ for the hadron spectra per trigger photon will be slightly enhanced at small $p_{\rm T}^{\gamma}$ by the CNM effects.
At very high $p_{\rm T}^{\gamma}$, the CNM effect has no \textcolor{black}{notable} influence on $\gamma$-hadron spectra in mid-rapidity \cite{Dai:2013xca,Ma:2018tjv} in both A + A and p + A collisions.

From the above numerical calculations, the effect of cold nuclear matter only leads to a slight enhancement of the $\gamma$-hadron spectra at intermediate $p_{\rm T}^{\gamma}<35$ GeV/$c$. The suppressions of $\gamma$-triggered hadron spectra should be mainly caused by parton energy loss if it is observed in A + A or in p + A collisions.

\begin{widetext}

\begin{figure}[tbh]
\begin{center}
\includegraphics[width=0.4\textwidth]{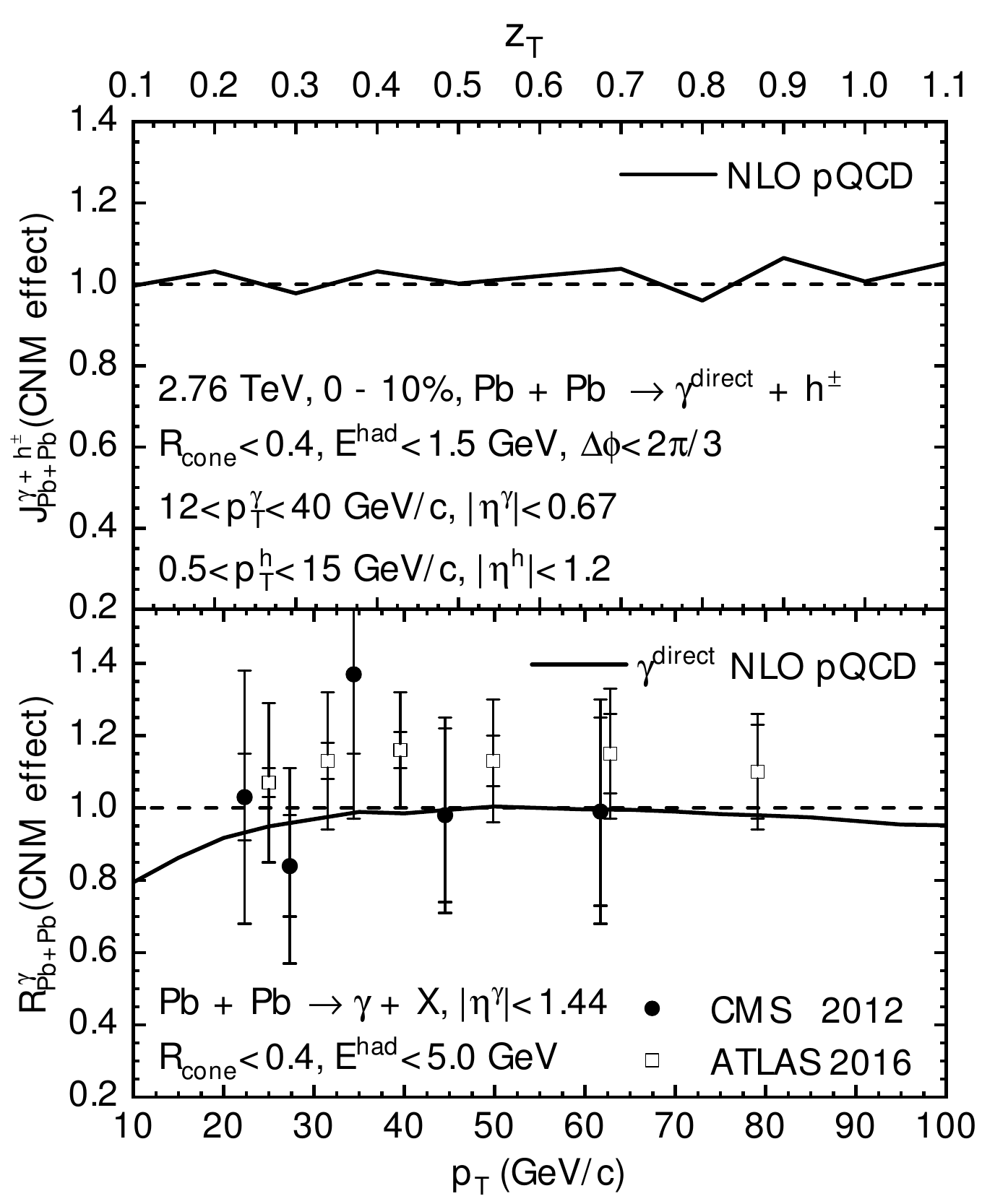}
\hspace{5mm}
\includegraphics[width=0.4\textwidth]{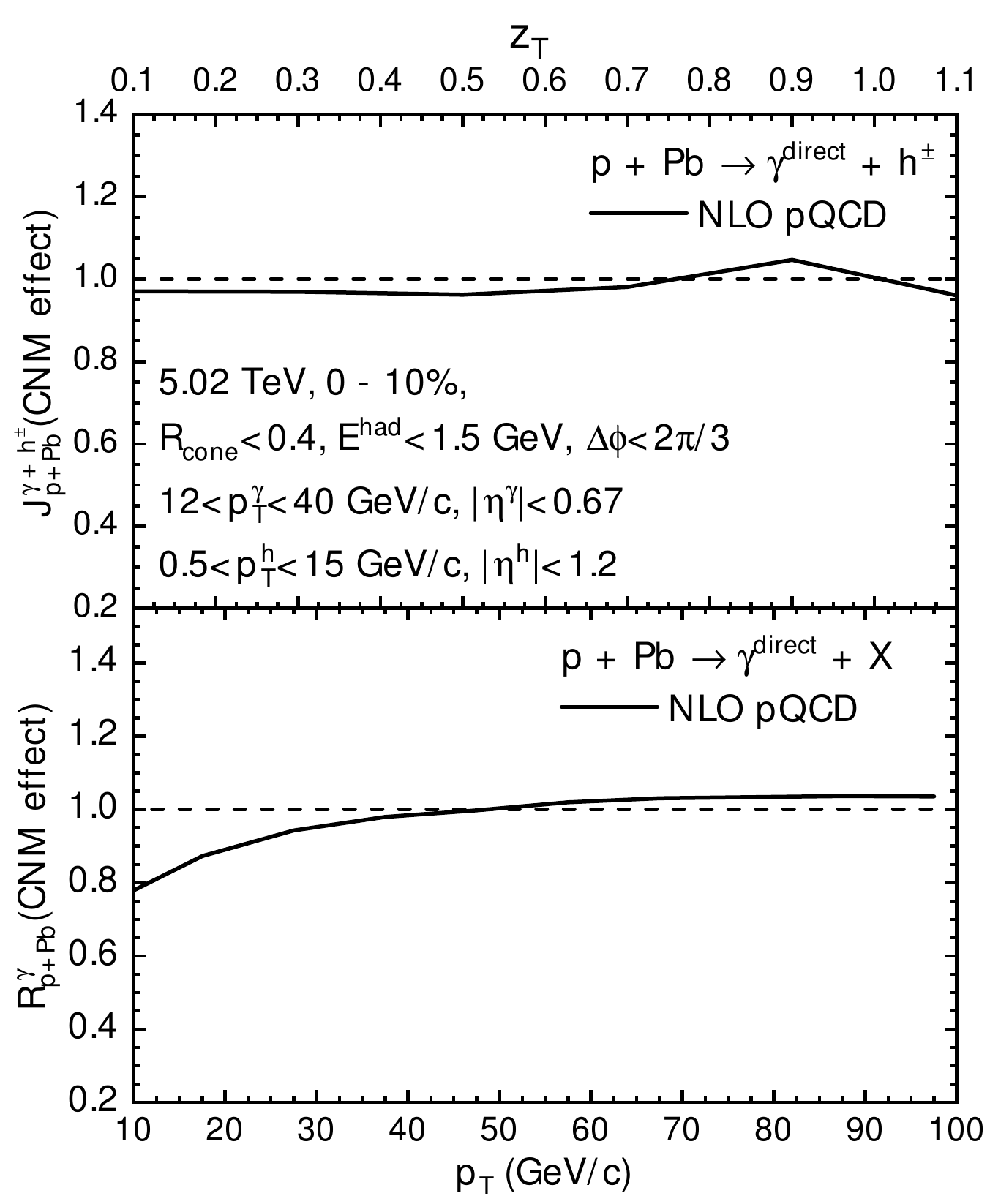}
\end{center}
\vspace{-5mm}
\caption[*]{Left:  The modification factor due to CNM effect on $\gamma^{\rm dir}$-hadron spectra with ($12<p_{\rm T}^{\gamma}<40$ GeV/$c$, $0.5<p_{\rm T}^{h}<15$ GeV/$c$) and on direct photon productions as compared with experimental data \cite{Chatrchyan:2012vq,Aad:2015lcb} in 0 - 10\% Pb + Pb collisions at $\sqrt{s_{\rm NN}}=2.76$ TeV. Right: The CNM modification factor on $\gamma^{\rm dir}$-hadron productions with ($12<p_{\rm T}^{\gamma}<40$ GeV/$c$, $0.5<p_{\rm T}^{h}<15$ GeV/$c$) and on direct photon productions in 0 - 10\% p + Pb collisions at $\sqrt{s_{\rm NN}}=5.02$ TeV. The $\gamma^{\rm dir}$-hadron suppression factors $J_{AA}^{\gamma h}(z_{\rm T})$ without normalization by the number of trigger photons is shown in the upper panels while the direct photon suppression factors $R_{AA}^{\gamma}(p_{\rm T}^{\gamma})$ is shown in the lower panels.}
\label{fig:CNM-pb-ppb}
\end{figure}
\begin{figure}[htb!]
\begin{center}
\includegraphics[width=0.4\textwidth]{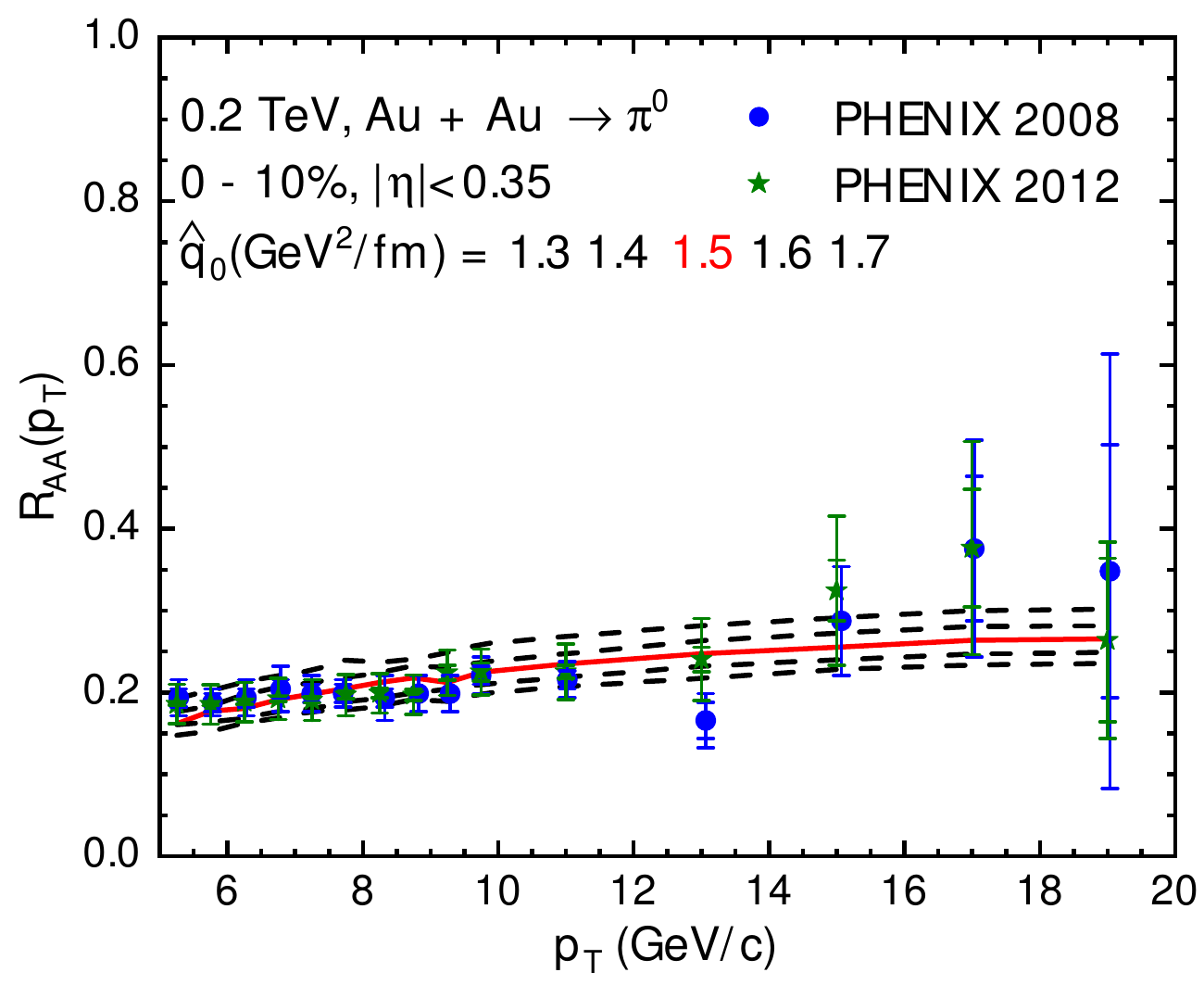}
\hspace{5mm}
\includegraphics[width=0.4\textwidth]{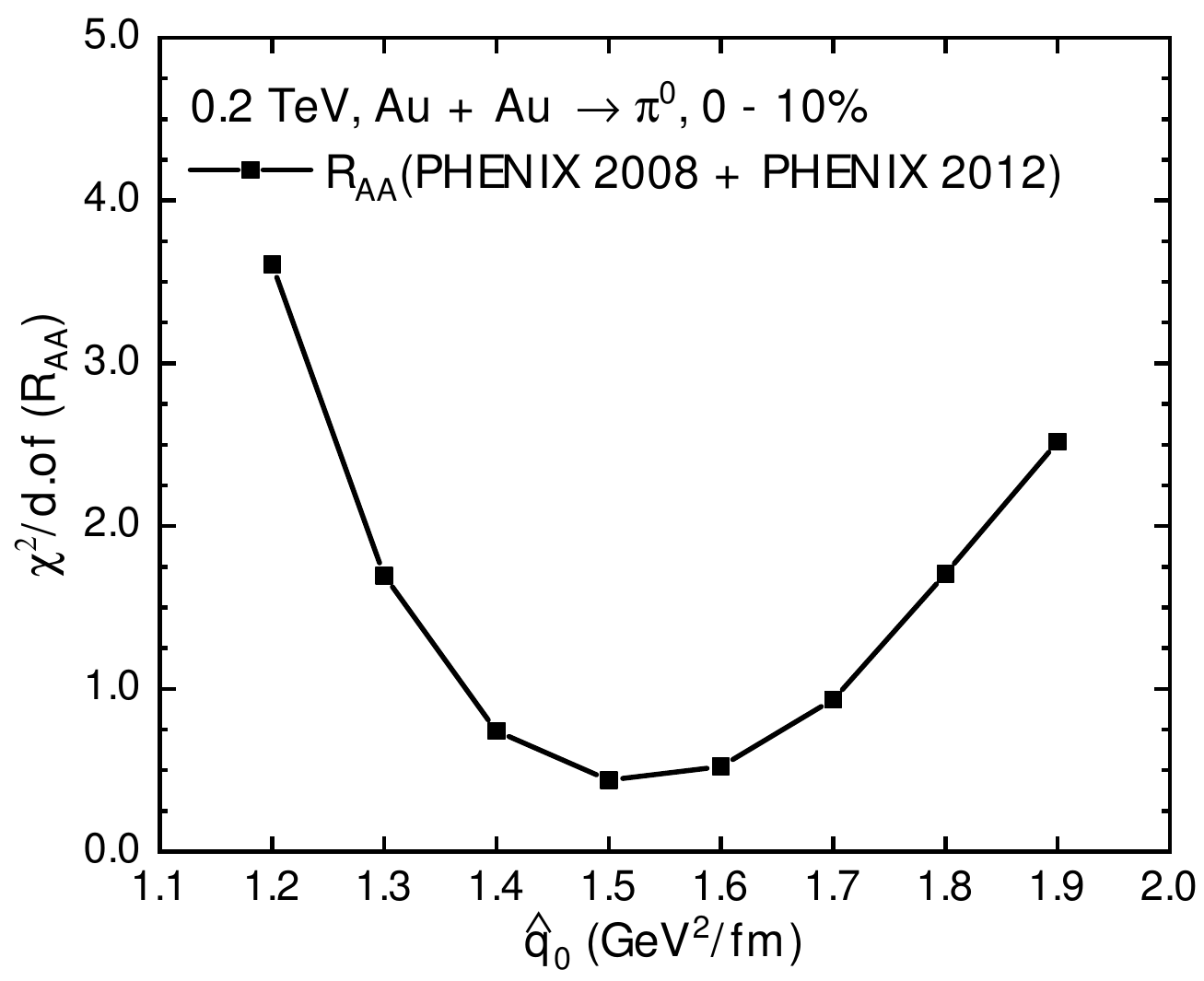}
\end{center}
\vspace{-5mm}
\caption{The single hadron suppression factor (left panel) in 0 - 10\% Au + Au collisions at $\sqrt{s_{\rm NN}}=0.2$ TeV compared with PHENIX \cite{Adare:2008qa,Adare:2012wg} data and the corresponding  $\chi^2/d.o.f$ of the fit as a function of the initial jet transport coefficient $\hat q_0$ (right panel).}
\label{fig:RAuAu}
\end{figure}
\begin{figure}[htb!]
\begin{center}
\includegraphics[width=0.4\textwidth]{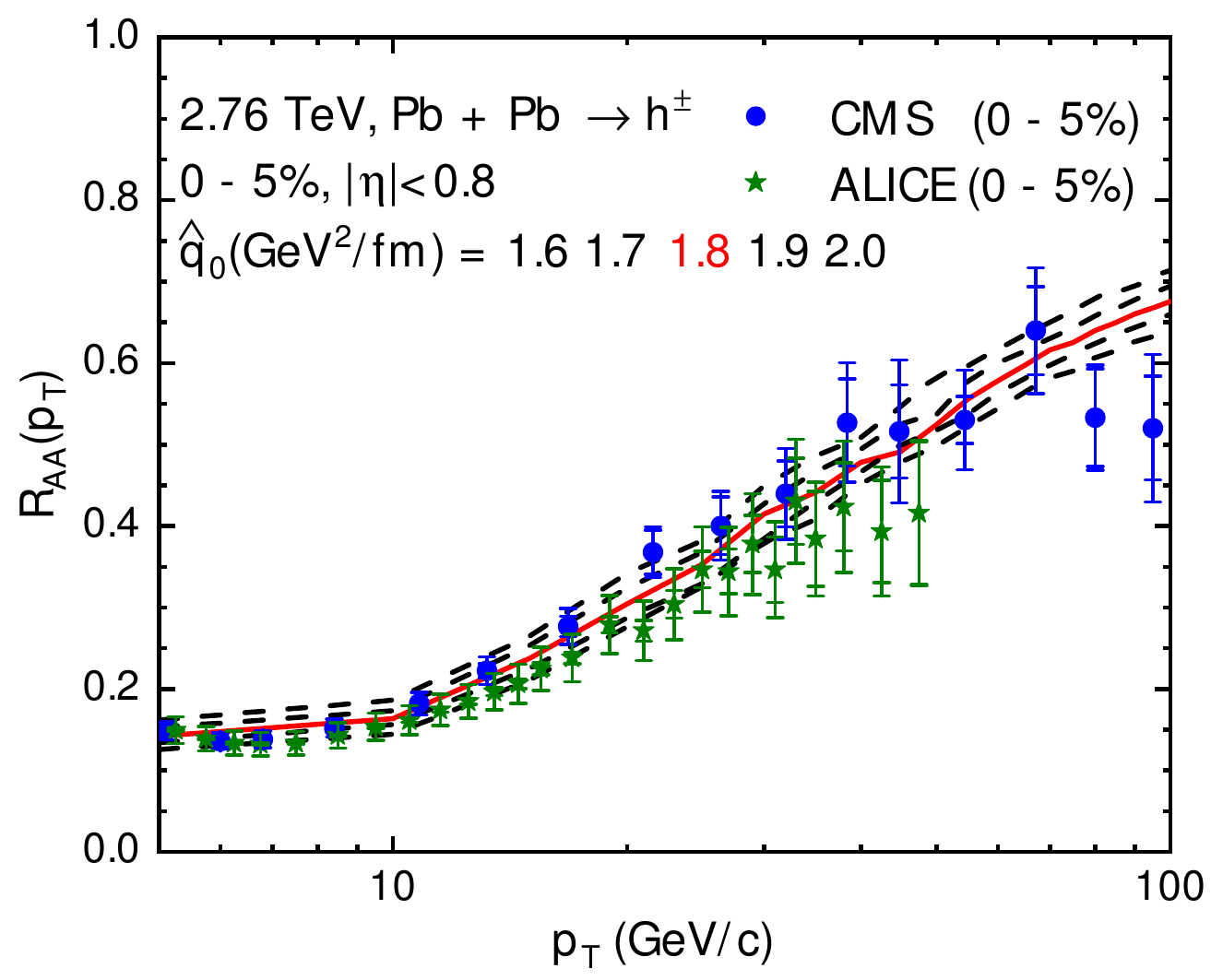}
\hspace{5mm}
\includegraphics[width=0.4\textwidth]{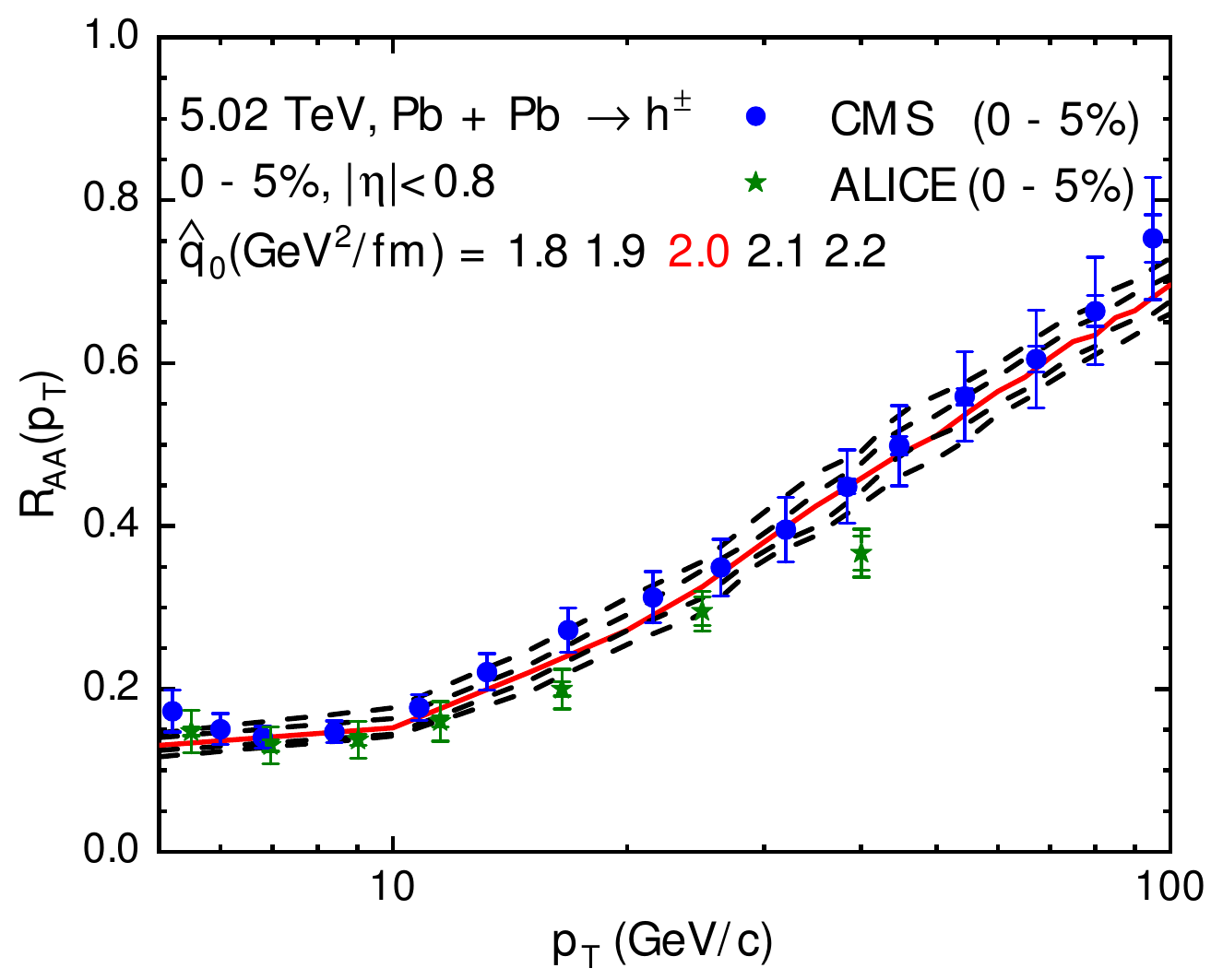}
\includegraphics[width=0.4\textwidth]{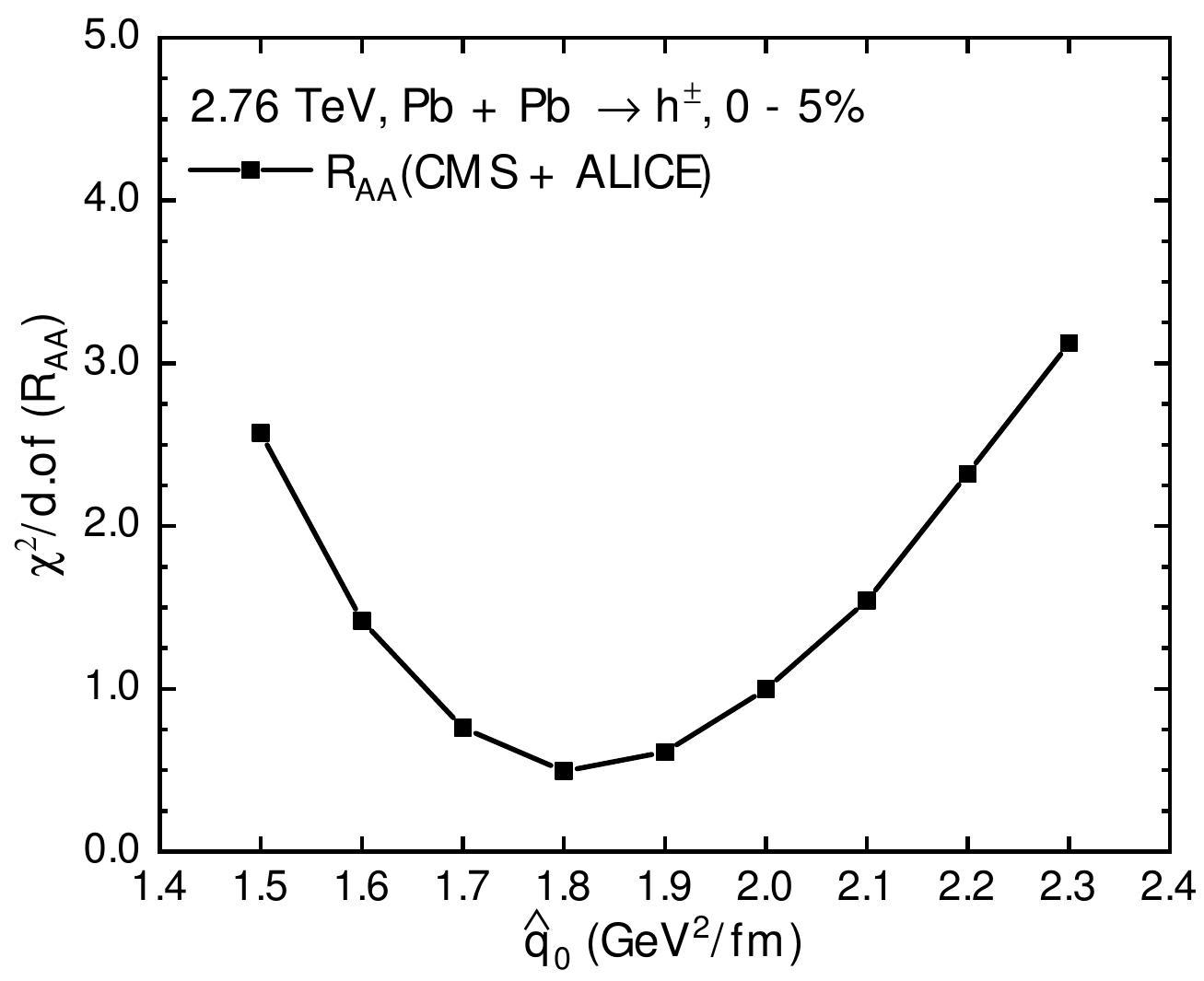}
\hspace{5mm}
\includegraphics[width=0.4\textwidth]{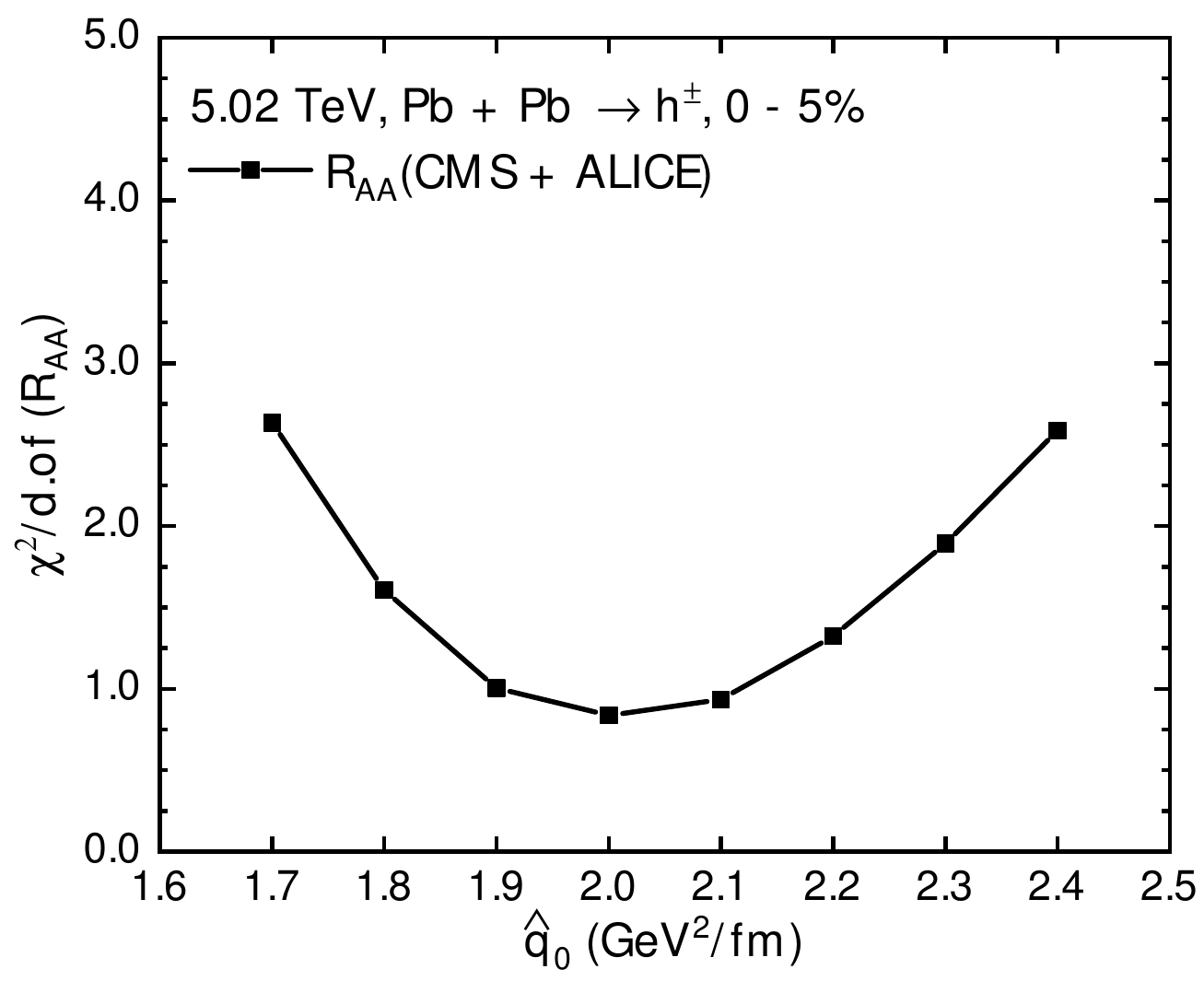}
\end{center}
\vspace{-5mm}
\caption[*]{The single hadron suppression factors (upper panels) in central 0 - 5\% Pb + Pb collisions at $\sqrt{s_{\rm NN}}=2.76$ TeV (left \textcolor{black}{panel})  and  $\sqrt{s_{\rm NN}}=5.02$ TeV (right \textcolor{black}{panel}) compared with CMS \cite{CMS:2012aa, Khachatryan:2016odn} and ALICE \cite{Abelev:2012hxa, Acharya:2018qsh} data and the corresponding $\chi^2/d.o.f$ of the fits as a function of the initial jet transport coefficient $\hat q_0$ (lower panels).}
\label{fig:RPbPb}
\end{figure}

\end{widetext}

\section{ Jet transport coefficient from suppression of single inclusive hadron spectra}

In order to describe jet quenching in high-energy heavy-ion collisions in the pQCD model, one needs to have the space-time evolution of the jet transport coefficient in Eq.~(\ref{eq:qhat}) along the parton propagation. The dynamical evolution of the QGP medium that governs the space-time evolution of the local temperature and flow velocity in our study of jet quenching in A + A collisions is obtained using the (2+1)-dimensional viscous hydrodynamic model VISH (2+1) with Monte-Carlo Glauber (MC-Glauber) initial conditions \cite{Song:2007fn,Song:2007ux,Qiu:2011hf,Qiu:2012uy}.

The scaled jet transport coefficient $\hat q/T^3$ in the co-moving frame in principle should also depend on the local temperature. The best way to extract such temperature-dependent $\hat q/T^3$ is to do a global fit using advanced inference technique such as Bayesian method with model emulations \cite{Soltz:2019aea}. Before such a comprehensive and expensive (in computing time) analysis, it is a common practice to assume a constant value of $\hat q/T^3$ for each centrality and colliding energy since most of the jet quenching comes from the early stage of the bulk evolution when the local temperature is the highest.
However, the extracted value of $\hat q/T^3$ can still depend on centrality and colliding energy due to its intrinsic temperature dependence.  Indeed, the initial  scaled  jet transport coefficient $\hat{q}_0/T_0^3$ extracted in a previous work \cite{Xie:2019oxg} from single and dihadron suppression in central Au + Au collisions at $\sqrt{s_{\rm NN}}=0.2$ TeV, Pb + Pb collisions at $\sqrt{s_{\rm NN}}=2.76$ and 5.02 TeV, and Xe + Xe collisions at $\sqrt{s_{\rm NN}}=5.44$ TeV is found to decrease slightly with the initial temperature $T_0$. This indicates a systematic error for  the value of \textcolor{black}{$\hat{q}_0/T_0^3$} if one assumes it a constant along the jet propagation path for a given centrality in A + A collisions at a given colliding energy.

To take into account these uncertainties, we extend the extractions of $\hat q_0/T_0^3$ from single hadron suppression to four different centralities,  0 - 5\%, 20 - 30\%, 40 - 50\% and 60 - 70\% in Au + Au collisions at $\sqrt{s_{\rm NN}}=0.2$ TeV, Pb + Pb collisions at $\sqrt{s_{\rm NN}}=2.76$ and 5.02 TeV.
As three examples of such extractions, we show in Figs.~\ref{fig:RAuAu} and \ref{fig:RPbPb} the $\chi^2$ fits to the suppression of single inclusive hadron spectra in 0 - 10\% central Au + Au collisions at 0.2 TeV, 0 - 5\% central  Pb + Pb collisions at 2.76 TeV and 5.02 TeV, respectively. The extracted values of the jet transport coefficient are  $\hat q_0=1.5$ GeV$^2/$fm or  $\hat q_0/T_0^3=5.5$ at $T_0=380$ MeV,  $\hat q_0=1.8$ GeV$^2/$fm or $\hat q_0 /T^3_0=3.1$ at $T_0=486$ MeV  and $\hat{q}_0=2.0 $ GeV$^2/$fm or $\hat q_0/T_0^3=2.9$  at $T_0=515 $ MeV, respectively.  We can see that the initial value of the scaled jet transport coefficient $\hat{q}_0/T_0^3$ in the center of the most central A + A collisions decreases with the initial temperature achieved at increasing colliding energy. The value of $\hat{q}_0$ in central Pb + Pb collisions at 5.02 TeV is only slightly larger than at 2.76 TeV, even though the charged hadron rapidity density is about 20\% higher at 5.02 TeV \cite{Acharya:2018qsh}.

We should note that our numerical calculations and extraction of the jet transport coefficient here are somewhat different from a previous work in Ref. \cite{Xie:2019oxg}.  Different parametrization of FFs are used there that cause non-negligible effect on the extracted values of $\hat{q}_0$. We have also made an improvement to the parton energy loss formula in Eq. (\ref{eq:deltaE}) in which we subtract $\tau_0$ from the total propagation time so that their difference enters the variable in $\sin^2\left[l_{\rm T}^{2}(\tau-\tau_0)/4z(1-z)E\right]$ due to LPM interference. As a result, the extracted values of $\hat{q}_0/T_0^3$ are a little larger than those in the previous work \cite{Xie:2019oxg}. However,  both values are consistent with that from the JET collaboration \cite{Burke:2013yra} within the uncertainty range.

\begin{figure}[htb!]
\vspace{3mm}
\begin{center}
\includegraphics[width=0.4\textwidth]{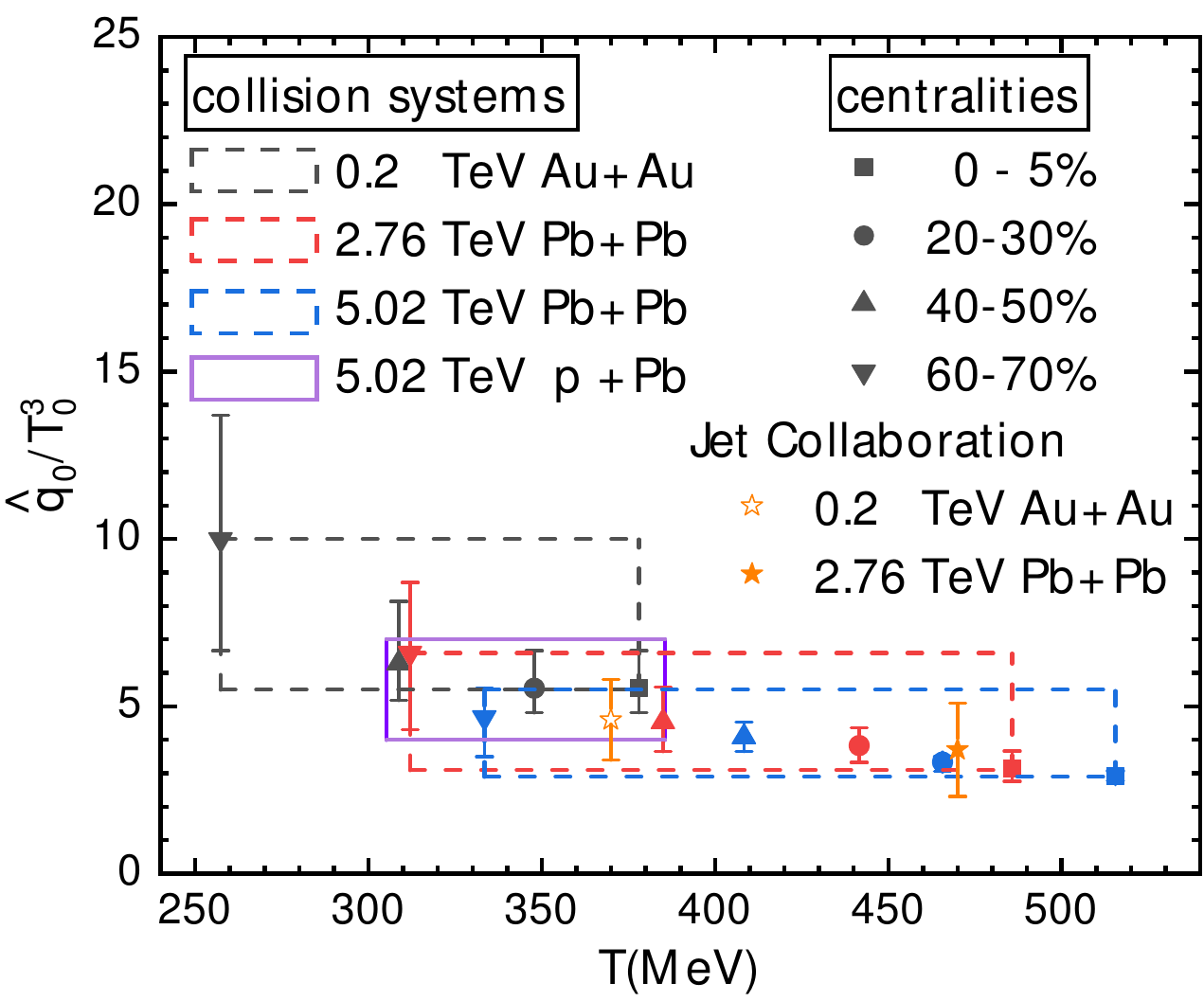}
\end{center}
\vspace{-5mm}
\caption{The scaled jet transport parameter \textcolor{black}{$\hat{q}_0/T_0^3$} as a function of the initial highest temperature in the center of the collision systems extracted from single hadron suppression in Au + Au at $\sqrt{s_{\rm NN}}=0.2$ TeV,  Pb + Pb collisions at $\sqrt{s_{\rm NN}}=2.76$ and 5.02 TeV in different (0 - 5\%, 20 - 30\%, 40 - 50\%, 60 - 70\%) centralities. The results for Au + Au at $\sqrt{s_{\rm NN}}=0.2$ TeV are denoted by black symbols; for Pb + Pb at $\sqrt{s_{\rm NN}}=2.76$ TeV by red symbols; for Pb + Pb collisions at $\sqrt{s_{\rm NN}}=5.02$ TeV by blue symbols. Symbols with same color from right to left represent the results from central to peripheral collisions. The two orange star points are the results from JET collaboration \cite{Burke:2013yra}. The rectangles indicate the range of $\hat{q}_0/T_0^3$ for each collision system at a given colliding energy.}
\label{fig:q-T3}
\end{figure}

The temperature dependence of the scaled jet transport coefficient \textcolor{black}{$\hat{q}_0/T_0^3$} from these extractions are summarized in Fig. \ref{fig:q-T3}. We observe a clear but small temperature dependence for the temperature range achieved in heavy-ion collisions at the LHC energies. The scaled jet transport coefficient $\hat q_0/T_0^3$ decreases slightly with \textcolor{black}{increasing} $T_0$ as also indicated by the values extracted by the JET Collaboration \cite{Burke:2013yra}. Such a weak temperature dependence is also observed in more recent study \cite{Feal:2019xfl}.

The temperature dependence of the $\hat{q}_0/T_0^3$ extracted from suppression of single inclusive hadron spectra in A + A collisions with different centralities as shown in Fig. \ref{fig:q-T3} will provide us improved estimates of the systematic uncertainties associated with the assumption of a constant \textcolor{black}{$\hat{q}_0/T_0^3$} through the jet propagation path in A + A collisions with a given centrality and at a given colliding energy. These uncertainties for A + A collisions at both RHIC and LHC energies are indicated by dashed boxes in Fig. \ref{fig:q-T3}.
\vspace{5mm}

\begin{figure}[tbh]
\begin{center}
\includegraphics[width=0.4\textwidth]{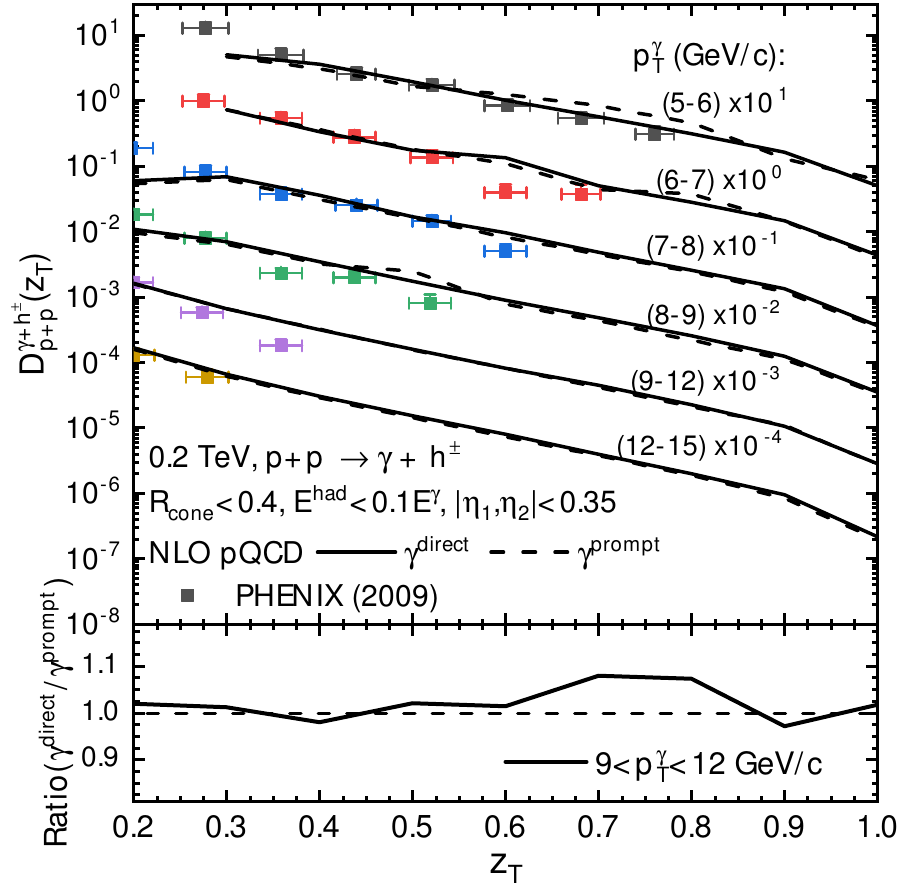}
\end{center}
\vspace{-5mm}
	\caption[*]{$\gamma^{\rm dir}$ ($\gamma^{\rm prompt}$)-triggered FFs with six different $p_{\rm T}^{\gamma}$ ranges in p + p collisions at $\sqrt{s_{\rm NN}}=0.2$ TeV \textcolor{black}{scaled by factors for better visualization and} compared with PHENIX data \cite{Frantz:2009zn}. \textcolor{black}{The ratio of contribution of direct photon triggered FFs to prompt photon triggered FFs with $9<p_{\rm T}^{\gamma}<12$ GeV$/c$ as an example is shown in lower panel.}}
\label{fig:Dpp-200}
\end{figure}
\section{$\gamma$-hadron spectra and jet quenching in A + A collisions}

In this section we will focus on the medium modification of $\gamma$-triggered hadron spectra in A + A collisions due to parton energy loss in hot QGP within the pQCD parton model \textcolor{black}{via the medium-modified parton fragmentation functions}. Shown in Fig. \ref{fig:Dpp-200} are the calculated $\gamma^{\rm dir}$ (solid lines) and  $\gamma^{\rm prompt}$-triggered (dashed lines) fragmentation functions in p + p collisions at $\sqrt{s_{\rm NN}}=0.2$ TeV as a baseline which agrees well with the PHENIX data \cite{Frantz:2009zn}.
These results are an updated version of the Fig. 1 in Ref.\cite{Zhang:2009rn}. The differences between them are negligible and come from the different choices of factorization scale and parton distribution functions. We use the factorization scale $\mu=1.2p_{\rm T}^{\gamma}$ in this study instead of $0.5p_{\rm T} ^{\gamma}$ in the previous study. We also use the updated parton distribution functions in a nucleon as given by the CT14 \cite{Hou:2016nqm} parameterizations instead of CTEQ6M parameterizations\cite{Stump:2003yu}. We note that the fragmentation functions triggered by $\gamma^{\rm prompt}$ are similar to that triggered by $\gamma^{\rm dir}$ with the isolation cuts. Therefore, we only focus on $\gamma^{\rm dir}$-hadron spectra in the following discussions.

\begin{figure}[tbh]
\begin{center}
\includegraphics[width=0.4\textwidth]{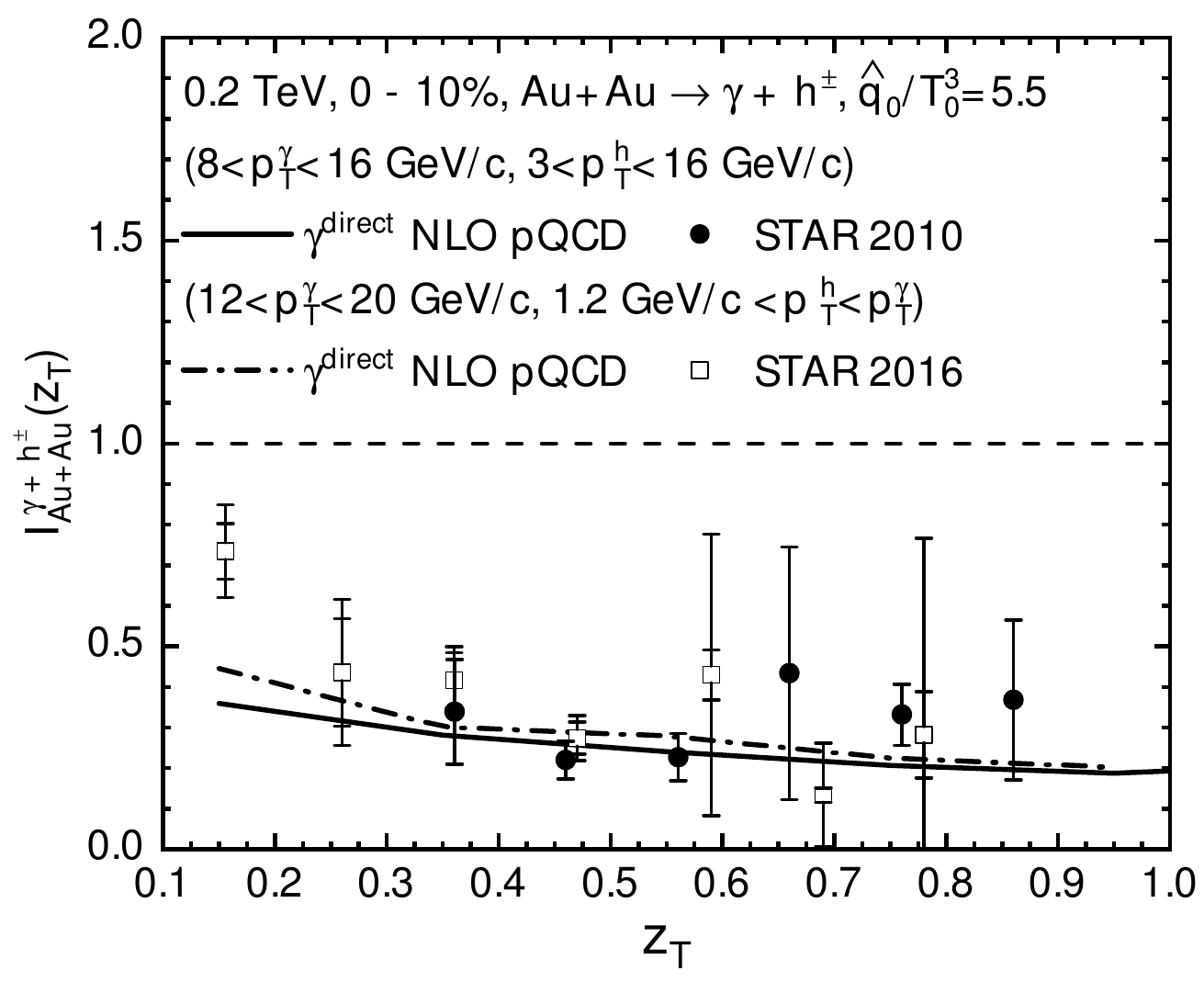}
\end{center}
\vspace{-5mm}
	\caption[*]{$\gamma^{\rm dir}$-hadron suppression factors with $\hat{q}_0/T_0^3=5.5$ in 0 - 10\% Au + Au collisions at $\sqrt{s_{\rm NN}}=0.2$ TeV with $8<p_{\rm T}^{\gamma}<16$ GeV/$c$, $3<p_{\rm T}^{h}<16$ GeV/$c$ (solid) and $12<p_{\rm T}^{\gamma}<20$ GeV/$c$, $1.2 ~\rm GeV/$c$<p_{\rm T}^{h}<p_{\rm T}^{\gamma}$ (dot-dashed)  as compared with STAR data \cite{Abelev:2009gu,STAR:2016jdz}.}
\label{fig:IAuAu}
\end{figure}
\begin{widetext}

\begin{figure}[tbh]
\begin{center}
\includegraphics[width=1.0\textwidth]{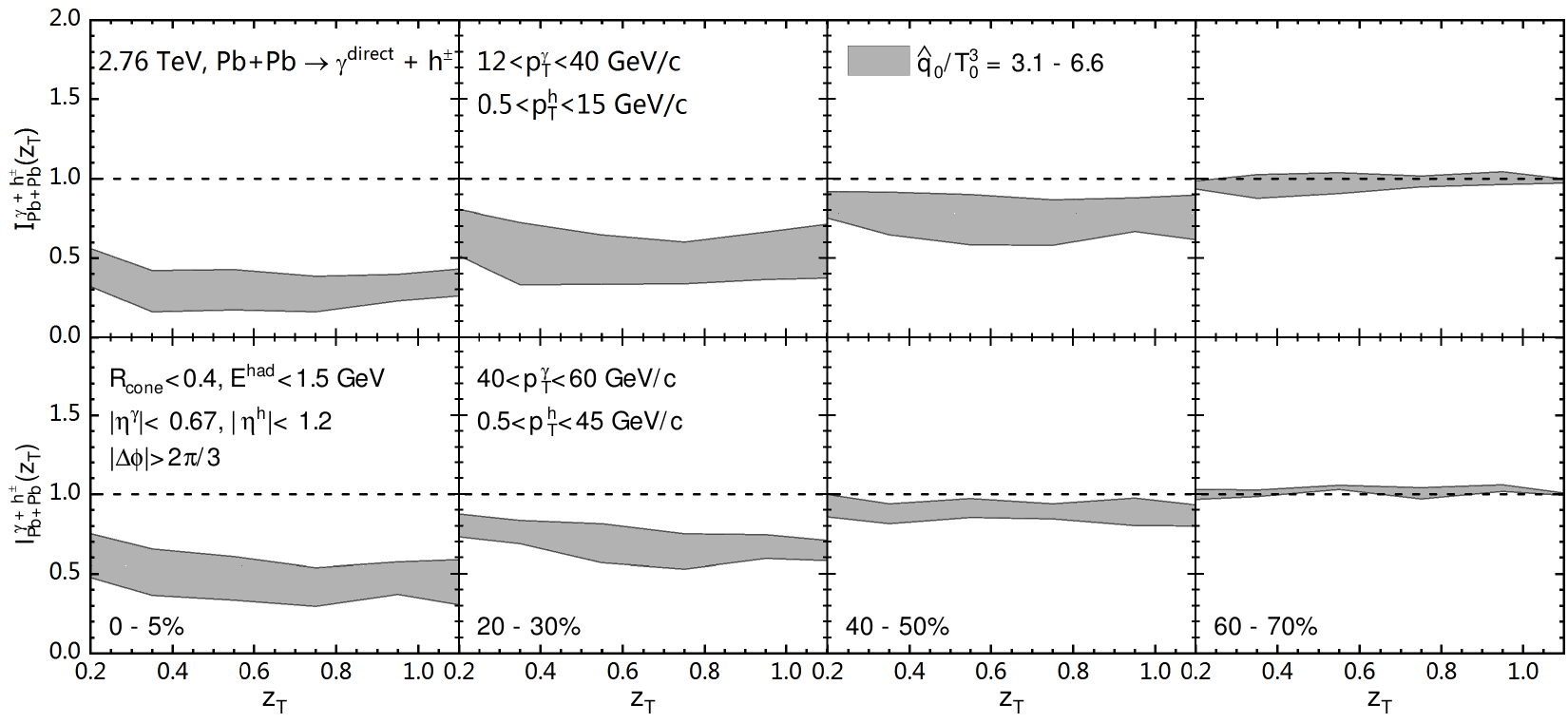}
\includegraphics[width=1.0\textwidth]{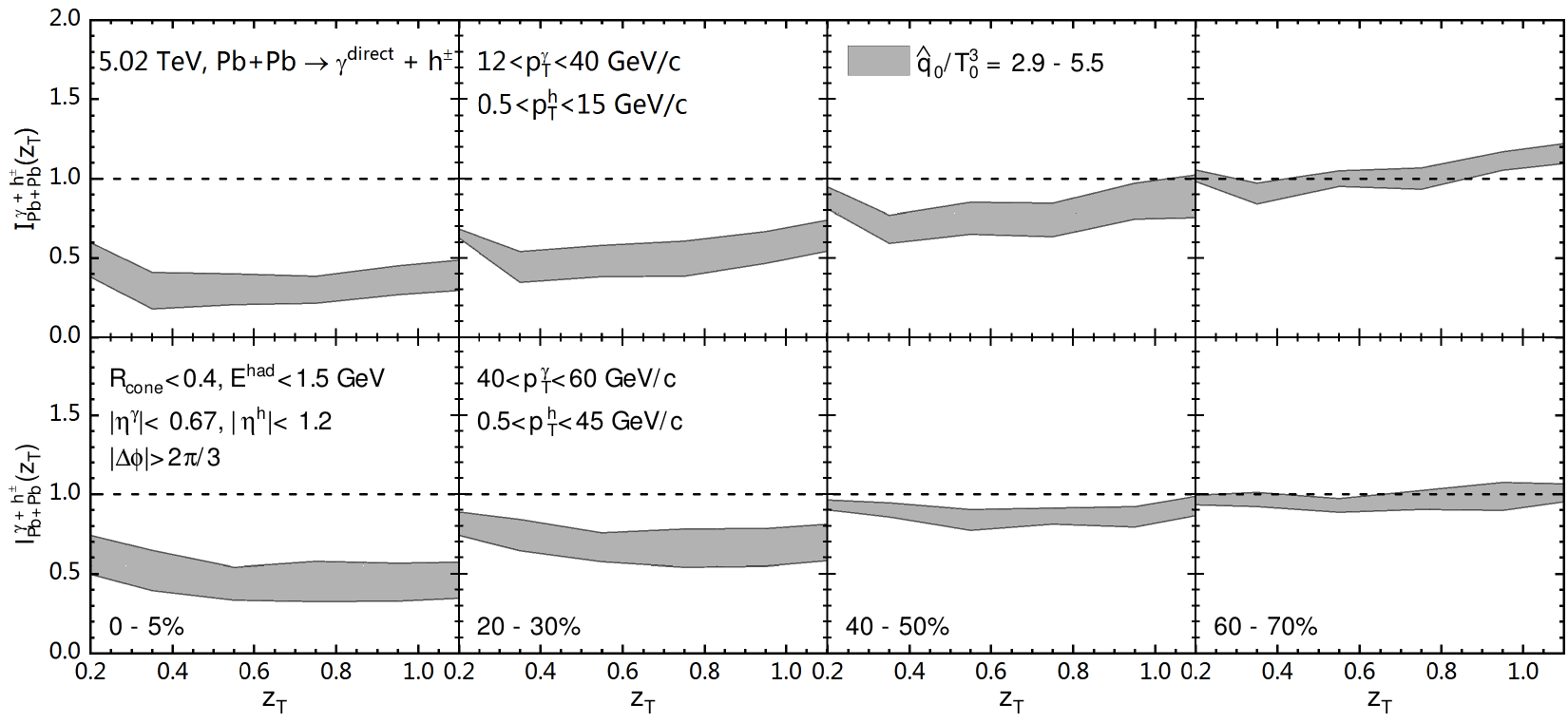}
\end{center}
\vspace{-5mm}
	\caption[*]{$\gamma^{\rm dir}$-hadron suppression factors as a function of $z_{\rm T}$ in 0 - 5\%, 20 - 30\%, 40 - 50\% and 60 - 70\%  Pb + Pb collisions, with $12<p_{\rm T}^{\gamma}<40$~GeV/$c$, $0.5<p_{\rm T}^{h}<15$~GeV/$c$ (upper panels)  and $40<p_{\rm T}^{\gamma}<60$~GeV/$c$, $0.5<p_{\rm T}^{h}<45$~GeV/$c$ (lower panels). The bands represent the range of  \textcolor{black}{$\hat{q}_0/T_0^3$}$=3.1- 6.6$ (upper figure) at $\sqrt{s_{\rm NN}}=2.76$ TeV and \textcolor{black}{$\hat{q}_0/T_0^3$}$=2.9 - 5.5$ (lower figure) at $\sqrt{s_{\rm NN}}=5.02$ TeV.}
\label{fig:Ipbpb}
\end{figure}

\end{widetext}

We first show the medium modification factor $I_{AuAu}^{\gamma h}$ for $\gamma$-triggered hadron spectra in 0 - 10\% Au + Au collisions at 0.2 TeV with  $8<p_{\rm T}^{\gamma}<16$ GeV/$c$, $3<p_{\rm T}^{h}<16$ GeV/$c$ (solid line) and $12<p_{\rm T}^{\gamma}<20$ GeV/$c$, $1.2 ~\rm GeV/$c$ <p_{\rm T}^{h}<p_{\rm T}^{\gamma}$ (dot-dashed line) as compared with STAR experimental data \cite{Abelev:2009gu,STAR:2016jdz} in Fig.~\ref{fig:IAuAu}.
In the pQCD model calculations the scaled initial jet transport coefficient $\hat{q}_0/T_0^3=5.5$ is used.

One can see that $\gamma$-triggered hadron spectra are suppressed by nearly 80\% due to jet quenching in central Au + Au collisions at $\sqrt{s_{\rm NN}}=0.2$ TeV.  Our results are consistent with the experimental data except the last data point at small $z_{\rm T}=0.15$ where contributions from hadronization of radiated gluons and jet-induced medium recoil partons \cite{Chen:2017zte} become important.

We note that the modification factor $I_{Au Au}^{\gamma h}$ as a function of $z_{\rm T}$ increases slightly with $p_{\rm T}^{\gamma}$ especially at intermediate and small $z_{\rm T}$. This is because the parton energy loss has an energy dependence that is weaker than a linear dependence \cite{Wang:2016fds}, so the fractional energy loss $\Delta E/E$ actually decreases with jet energy. The fraction of punch-through jets that come out and fragment into hadrons without energy loss also increases with $p_{\rm T}^{\gamma}$ and leads to increase of $I_{Au Au}^{\gamma h}$.

Using the ranges of the scaled initial jet transport coefficient $\hat{q}_0/T_0^3$ as extracted from single inclusive hadron spectra in the \textcolor{black}{previous} section, $\hat{q}_0/T_0^3=3.1 - 6.6$ at $\sqrt{s_{\rm NN}}=2.76$ TeV and $\hat{q}_0/T_0^3=2.9 - 5.5$ at $\sqrt{s_{\rm NN}}=5.02$ TeV, we can also predict the medium modification factors for $\gamma$-triggered hadron spectra in Pb + Pb collisions at both colliding energies for different (0 - 5\%, 20 - 30\%, 40 - 50\%, 60 - 70\%) centralities as shown in Fig. \ref{fig:Ipbpb}.
Two different ranges of $p_{\rm T}^{\gamma}$ and $p_{\rm T}^{h}$ are used: $12<p_{\rm T}^{\gamma}<40$~GeV/$c$, $0.5<p_{\rm T}^{h}<15$~GeV/$c$ for the results in the upper panels and $40<p_{\rm T}^{\gamma}<60$~GeV/$c$, $0.5<p_{\rm T}^{h}<45$~GeV/$c$ in the lower panels.

From the first plot of the upper figure in Fig. \ref{fig:Ipbpb} we see that $I_{PbPb}^{\gamma h}$ is about $0.2 \sim 0.4$ in 0 - 5\% central Pb + Pb collisions at 2.76 TeV and it increases with centrality.
In 60 - 70\% peripheral collisions, there is almost no suppression of $\gamma$-triggered hadron spectra. Similarly as at the RHIC energy, the suppression of $\gamma$-triggered hadron spectra becomes weaker at larger $p_{\rm T}^{\gamma}$. The results of $\gamma$-triggered hadron suppression at 5.02 TeV are almost the same as at 2.76 TeV, similar to the situation for single charged hadron suppression \cite{Khachatryan:2016odn, Acharya:2018qsh}.

\section{$\gamma$-triggered hadron spectra in  p + Pb  collisions}

In order to predict $\gamma$-triggered hadron spectra in p + Pb collisions in our pQCD model under the assumption that a small droplet of QGP is formed, one also needs to provide the space-time evolution of the QGP droplet and the value of the scaled initial jet transport coefficient \textcolor{black}{$\hat{q}_0/T_0^3$}.

We will use superSONIC (2+1) D hydrodynamic model \cite{Romatschke:2015gxa,Weller:2017tsr,Romatschke:2007mq,Luzum:2008cw} to describe the space-time evolution of the QGP droplet in p + Pb collisions. This model  gives similar results as the VISH (2+1) D model on the transverse dynamics of the bulk medium in A + A collisions with the same initial conditions.
\textcolor{black}{For example, with superSONIC (2+1) D hydrodynamic model the extracted jet transport parameter $\hat{q}_0$ in 0 - 10\% Au + Au collisions at 0.2 TeV, shown in Fig. ~\ref{fig:RHIC-super} of Appendix B, is about the same within the uncertainty (purple box in Fig.~\ref{fig:q-T3}) as that extracted with the VISH hydro as shown in Fig. \ref{fig:RAuAu}.}
The challenge for a hydrodynamic model in p + A collisions is the modeling of the fluctuating initial conditions. The model for initial conditions in superSONIC is tuned to describe p + A collisions and other small system as in d + Au and He + Au collisions \cite{Nagle:2013lja}. In principle, one should use a (3+1) D hydrodynamic model for p + A collisions since there is no longer the Bjorken scaling in the longitudinal direction. Since we restrict our study to jet quenching in the central rapidity region within a small rapidity window, the effect of the longitudinal dynamics due to the breaking of Bjorken scaling should be small on jet quenching observables averaged in a rapidity window centered at $y=0$, especially if the (2+1) D hydro and the initial conditions are tuned to fit the bulk hadron spectra and anisotropic flows.
\textcolor{black}{Nevertheless, we should note that to minimize systematic uncertainties from (2+1)D hydro models for a small asymmetric system, one should use a more realistic (3+1) D hydro \cite{Schenke:2010nt, Schenke:2020mbo} for p + A collisions in the future studies.}

\begin{figure}[htb!]
\begin{center}
\includegraphics[width=0.4\textwidth]{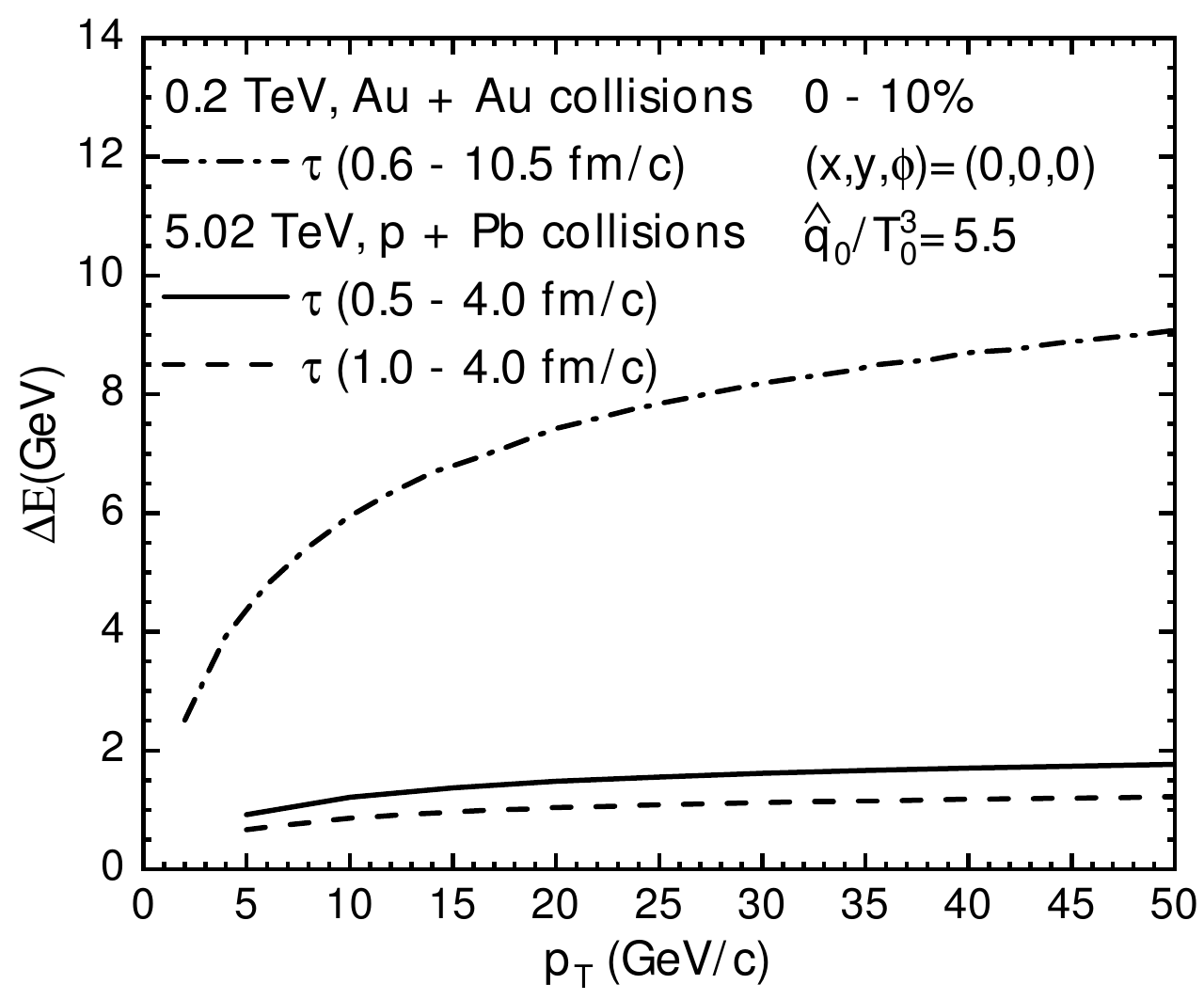}
\end{center}
\vspace{-5mm}
\caption{The energy loss of a light quark produced at ($x=y=0$) as a function of $p_{\rm T}$ in 0 - 10\% p + Pb collisions at 5.02 TeV with initial time $\tau_0=0.5$ (solid line) and 1.0 fm/$c$ (dashed line) compared with the quark energy loss in 0 - 10\% Au + Au collisions at 0.2 TeV with initial time $\tau_0=0.6$ fm/$c$ (dot-dashed line), both with $\hat{q}_0/T_0^3=5.5$.}
\label{fig:De}
\end{figure}

From the superSONIC hydrodynamic model, the initial highest temperature at the center of p + Pb collisions at $\sqrt{s_{\rm{NN}}}=5.02$ TeV fluctuates from event to event in the range of $300 - 385$ MeV.  According to the temperature dependence of $\hat q_0/T^3_0$ extracted from single hadron suppression in A + A collisions, we will consider a constant $\hat q_0/T^3_0$ in the range of $4.0 - 7.0$ in p + Pb collisions as indicated by the purple solid box in Fig.~\ref{fig:q-T3}. This range of temperature happens to overlap with that in central Au + Au collisions at $\sqrt{s_{\rm{NN}}}=0.2$ TeV, which, however, have a much larger system size and longer lifetime. Shown in Fig. \ref{fig:De}, are the average radiative parton energy loss of a light quark originating from the center of the hot medium as a function of its initial transverse momentum with  $\hat{q}_0/T_0^3=5.5$ in both central 0 - 10\% Au + Au collisions at 0.2 TeV (dot-dashed) and p + Pb collisions at 5.02 TeV (solid). The quark energy loss in central Au + Au collisions is more than a factor of 4 larger than that in p + Pb collisions with similar initial temperature due to the larger system size and longer lifetime of the QGP medium in central Au + Au collisions.
The energy loss of a gluon is simply 9/4 that of a light quark. Based on this, one expects that the suppression of $\gamma$-triggered hadron spectra in p + Pb collisions is significantly smaller than that in A + A collisions.

Additionally, to study the sensitivity of the radiative parton energy loss on the initial time $\tau_0$, we also vary $\tau_0$ in the calculation when a parton starts interacting with the hot medium and losing energy. With $\hat{q}_0/T_0^3=5.5$, we set the default (solid line) initial time $\tau_0=0.5$ fm/$c$ in p + Pb collisions  as provided by the superSONIC hydrodynamic model.  If we set $\tau_0=1.0$ fm/$c$ and therefore let the quark to start losing energy a little later, the average energy loss for the quark (dashed) is about  30\% smaller. We will consider such a variation of the initial time as a part of the systematic errors in the prediction of the $\gamma$-hadron spectra in p + Pb collisions.

\begin{widetext}

\begin{figure}[tbh]
\begin{center}
\includegraphics[width=1.0\textwidth]{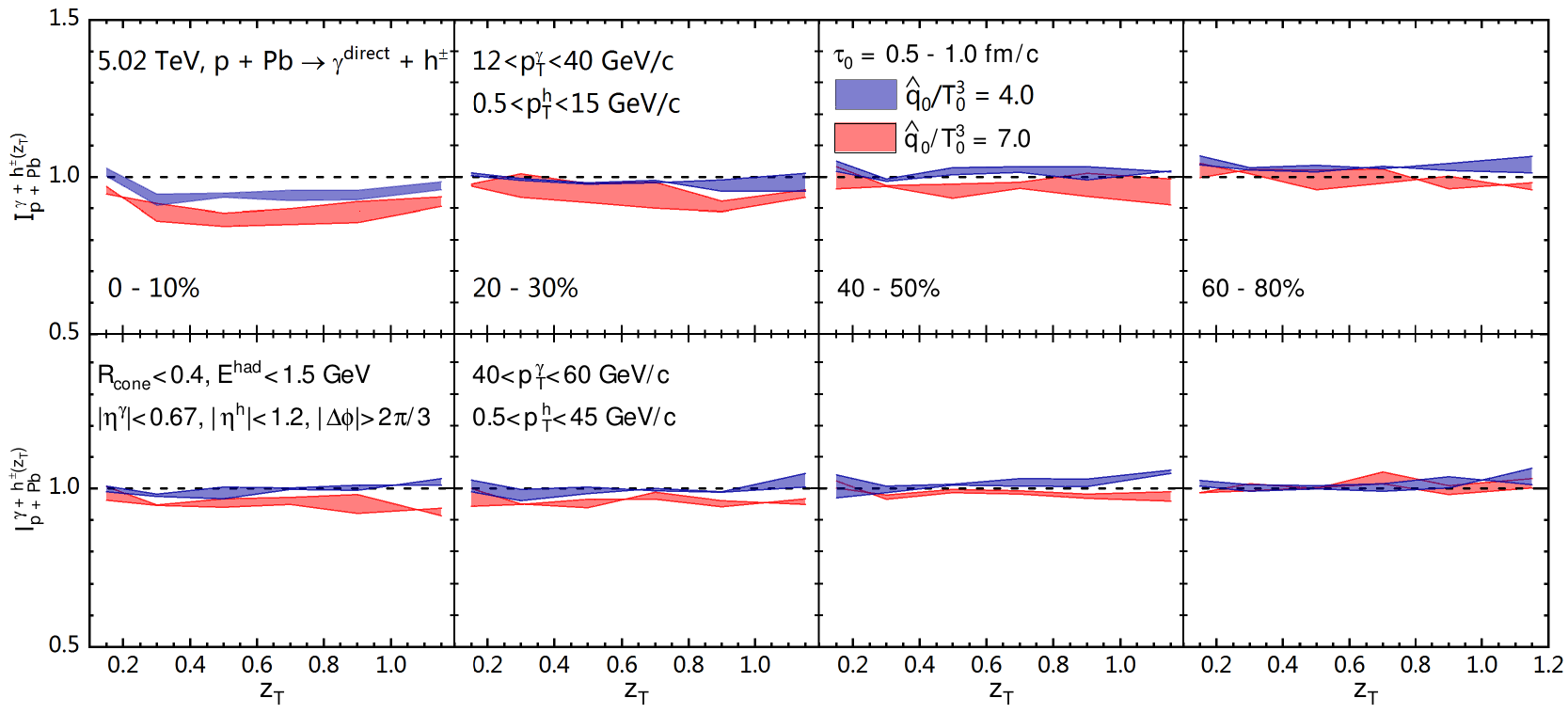}
\end{center}
\vspace{-5mm}
	\caption[*]{$\gamma^{\rm dir}$-hadron suppression factors as a function of $z_{\rm T}$ in 0 - 10\%, 20 - 30\%, 40 - 50\% and 60 - 80\% p + Pb collisions at $\sqrt{s_{\rm NN}}=5.02$ TeV with $12<p_{\rm T}^{\gamma}<40$~GeV/$c$, $0.5<p_{\rm T}^{h}<15$~GeV/$c$ (upper panels) and  $40<p_{\rm T}^{\gamma}<60$~GeV/$c$, $0.5<p_{\rm T}^{h}<45$~GeV/$c$ (lower panels) with $\hat{q}_0/T_0^3=4.0$ (blue bands) and 7.0 (red bands). The lower and  upper limit of bands indicate variations of the initial time between $\tau_0=0.5$ (lower) and 1.0 fm/$c$ (upper).}
\label{fig:Ippb-5020}
\end{figure}

\end{widetext}

To test the expectation of $\gamma$-hadron suppression due to jet quenching in a small droplet of QGP in p + A collisions, we show in Fig.~\ref{fig:Ippb-5020} the predictions of the suppression factor for $\gamma$-triggered hadron spectra in 5.02 TeV p + Pb collisions, \textcolor{black}{with two values of $\hat{q}_0/T^3_0$ representing the uncertainties on jet transport coefficient in the range of initial temperatures according to the hydrodynamic model as shown in Fig.~\ref{fig:q-T3}.} The predictions are provided for four different centralities and for two different ranges of the transverse momentum of the trigger photon and associated hadron.
For $12<p_{\rm T}^{\gamma}<40$~GeV/$c$, $0.5<p_{\rm T}^{h}<15$~GeV/$c$ in the upper panel, the $\gamma$-triggered hadron spectrum in the most 0 - 10\% central p + Pb collisions is suppressed by about 10 $-$ 15\% with $\hat{q}_0/T_0^3=7.0$ shown by the red shaded band and by about 5\% with $\hat{q}_0/T_0^3=4.0$ shown by the blue shaded band.
The suppression becomes smaller in more peripheral collisions and disappears in the most 60 - 80\% peripheral collisions. The shaded bands in these results indicate variations of the results when one varies the initial time for parton energy loss between $\tau_0=0.5$ and 1.0 fm/$c$\textcolor{black}{, which show a 5\% difference for the suppression of the $\gamma$-triggered hadron.}

For a large transverse momentum of the triggered photon, $40<p_{\rm T}^{\gamma}<60$~GeV/$c$, $0.5<p_{\rm T}^{h}<45$~GeV/$c$ (the lower panels), the $\gamma$-triggered hadron spectrum is suppressed by 5\% at most in the most central p + Pb collisions. The effect of varying initial time from $\tau_0=0.5$ to 1.0 fm/$c$ on the suppression factor is almost indistinguishable for this large $p_{\rm T}^{\gamma}$.

As we have expected, the suppression of $\gamma$-triggered hadron spectra in p + Pb collisions at $\sqrt{s_{\rm NN}}=5.02$ TeV is much smaller than that in A + A collisions even when similar highest initial temperature is reached in the center of the QGP medium in both systems. We predict a suppression of about 5 $-$ 15\% for $\gamma$-hadron spectra with $12 < p_{\rm T}^{\gamma}<40$ GeV/$c$  in 0 - 10\% central p + Pb collisions due to jet quenching if we assume a QGP droplet is formed and can be described by hydrodynamic evolution. The suppression becomes weaker with increasing $p_{\rm T}^{\gamma}$ and in more peripheral collisions.

\section{Summary and Discussions}

In this paper,  we study the suppression of $\gamma$-triggered hadron spectra in p + Pb collisions at $\sqrt{s_{\rm NN}}=5.02$ TeV within NLO perturbative QCD parton model with medium modified fragmentation function due to parton energy loss under the assumption that a QGP droplet is produced  and its evolution can be described by hydrodynamics. The evolution of the QGP medium and its temperature profile \textcolor{black}{in} p + Pb collisions is simulated event-by-event \textcolor{black}{by} using the superSONIC model, while the parton energy loss is calculated within the high-twist formalism.

We have taken into account and illustrated the CNM effect on $\gamma$-hadron spectra (hadron yield per trigger) which is negligible and the net suppression of $\gamma$-hadron spectra, if any, should be  caused mainly by parton energy loss. We predict that $\gamma$-triggered hadron spectra are suppressed due to jet quenching by about 5 $-$ 15\% for $12<p_{\rm T}^{\gamma}<40$ GeV/$c$ in the most 0 - 10\% central p + Pb collisions at 5.02 TeV, with the initial jet transport coefficient $\hat q_0/T_0^3$ extracted from the experimental data on the suppression of single inclusive hadron spectra in A + A collisions. The suppression is shown to decrease with increasing $p_{\rm T}^{\gamma}$ and in more peripheral collisions. We also provided predictions of $\gamma$-hadron suppression in Pb + Pb collisions at $\sqrt{s_{\rm NN}}=2.76$ and 5.02 TeV which are similar because of similar values of $\hat q_0/T_0^3$ as extracted from the suppression of single inclusive hadron spectra in Pb + Pb collisions at these two energies. The experimental measurements of such suppression could provide much stringent constraints on the formation and dynamic evolution of QGP droplets in p + A collisions.

Most of the systematic uncertainties of our predictions within the parton energy loss model come from the assumption of a constant scaled jet transport coefficient $\hat q/T^3$ which is shown to have a weak but non-negligible temperature dependence. Such uncertainties can be reduced in the future by assuming a generic temperature dependence of $\hat q/T^3$ in the calculation and constrained global fits using advanced inference technique such as Bayesian method.

\vspace{5mm}
\section*{ACKNOWLEDGMENTS}

We would like to thank Jamie Nagle, Jeffrey Ouellette and Paul Romatschke for providing the superSONIC hydro profiles of p + Pb collisions used in this study.
This work is supported by National Natural Science Foundation of China under grant Nos. 11935007, 11221504 and 11890714, the Director, Office of Energy Research, Office of High Energy and Nuclear Physics, Division of Nuclear Physics, of the U.S. Department of Energy under grant No. DE-AC02-05CH11231, the National Science Foundation (NSF) under grant No. ACI-1550228 within the framework of the JETSCAPE Collaboration.

\textcolor{black}{
\section*{Appendix: A}
The  cold nuclear modification factors for $\gamma^{dir}$-hadron $I_{AA}^{\gamma h}$ and $I_{pPb}^{\gamma h}$ normalized by the number of trigger photons as a function of $z_{\rm T}$ in Au + Au collisions at $\sqrt{s_{\rm NN}}=0.2$ TeV (left plot), Pb + Pb collisions at $\sqrt{s_{\rm NN}}=2.76$ TeV (middle plot) and p + Pb collisions at $\sqrt{s_{\rm NN}}=5.02$ TeV (right plot) all within 0 - 10\% centrality are shown in Fig.~\ref{fig:IpA-AA-CNM}  in correspondence with  Fig.~\ref{fig:CNM-AuAu} and Fig. \ref{fig:CNM-pb-ppb}. In 0 - 10\% Au + Au collisions at 0.2 TeV, at an average value of photon trigger transverse momentum $p_{\rm T}=12$ GeV/$c$, the direct photon spectrum has a suppression of about 10\% as shown in Fig.~\ref{fig:CNM-AuAu}.  While in 0 - 10\% Pb + Pb at 2.76 TeV  and p + Pb collisions at 5.02 TeV, at an average value of photon trigger transverse momentum $p_{\rm T}=26$ GeV/$c$, the direct photon spectra both have a suppression of about 10\% as shown in Fig.~\ref{fig:CNM-pb-ppb}. All of them lead to an enhancement of about 10\% to the $\gamma$-hadron modification factor $I^{\gamma h}$ as shown in Fig.~\ref{fig:IpA-AA-CNM}.
}

\textcolor{black}{
\section*{Appendix: B}
To show similarity of our numerical results with VISH (2+1) D and superSONIC (2+1) D hydrodynamic model for the bulk medium evolution,  we extract the jet transport parameter $\hat{q}_0$ with superSONIC (2+1) D hydrodynamic model in 0 - 10\% Au + Au collisions at $\sqrt{s_{\rm NN}} =0.2$ TeV shown in Fig.~\ref{fig:RHIC-super}, as a comparison to Fig. \ref{fig:RAuAu} which is obtained with VHIS (2+1) D hydrodynamic model. As we can see the best fit of $\hat{q}_0$ we get with superSONIC hydro in 0 - 10\% Au + Au collisions at 0.2 TeV is 1.7 GeV$^2/$fm ($\hat q_0/T_0^3\approx 6.7$) while it is 1.5 GeV$^2/$fm ($\hat q_0/T_0^3\approx 5.9$) with VISH hydro. These two values of $\hat q_0/T_0^3$ are in the extracted range of $\hat q_0/T_0^3\approx 4.0 - 7.0$ as shown in the purple solid box in Fig. \ref{fig:q-T3} which is the jet transport coefficient range we used for p + Pb collisions in Fig. \ref{fig:Ippb-5020}.}

\begin{widetext}

\begin{figure}[tbh]
\begin{center}
\includegraphics[width=0.32\textwidth]{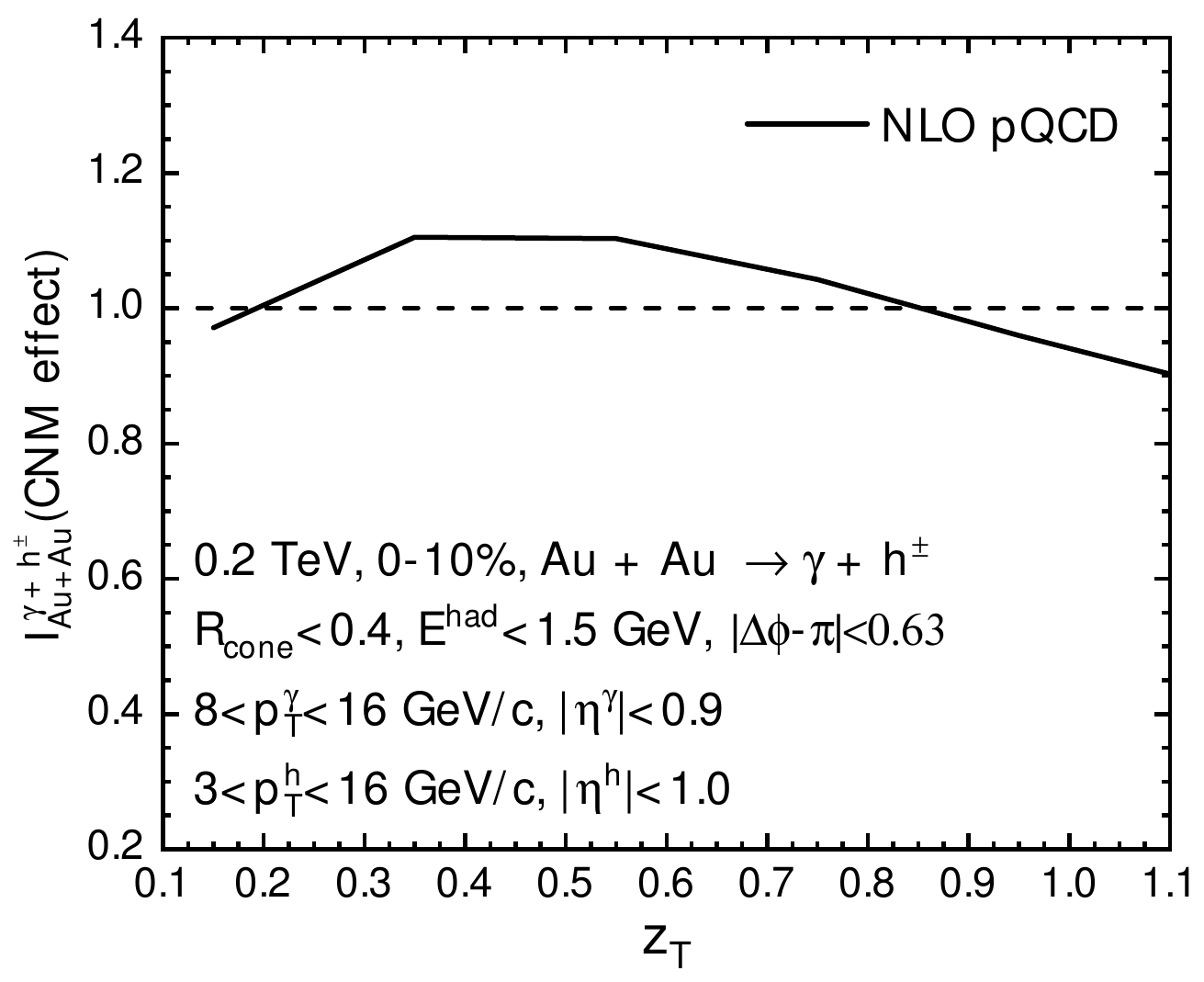}
\includegraphics[width=0.32\textwidth]{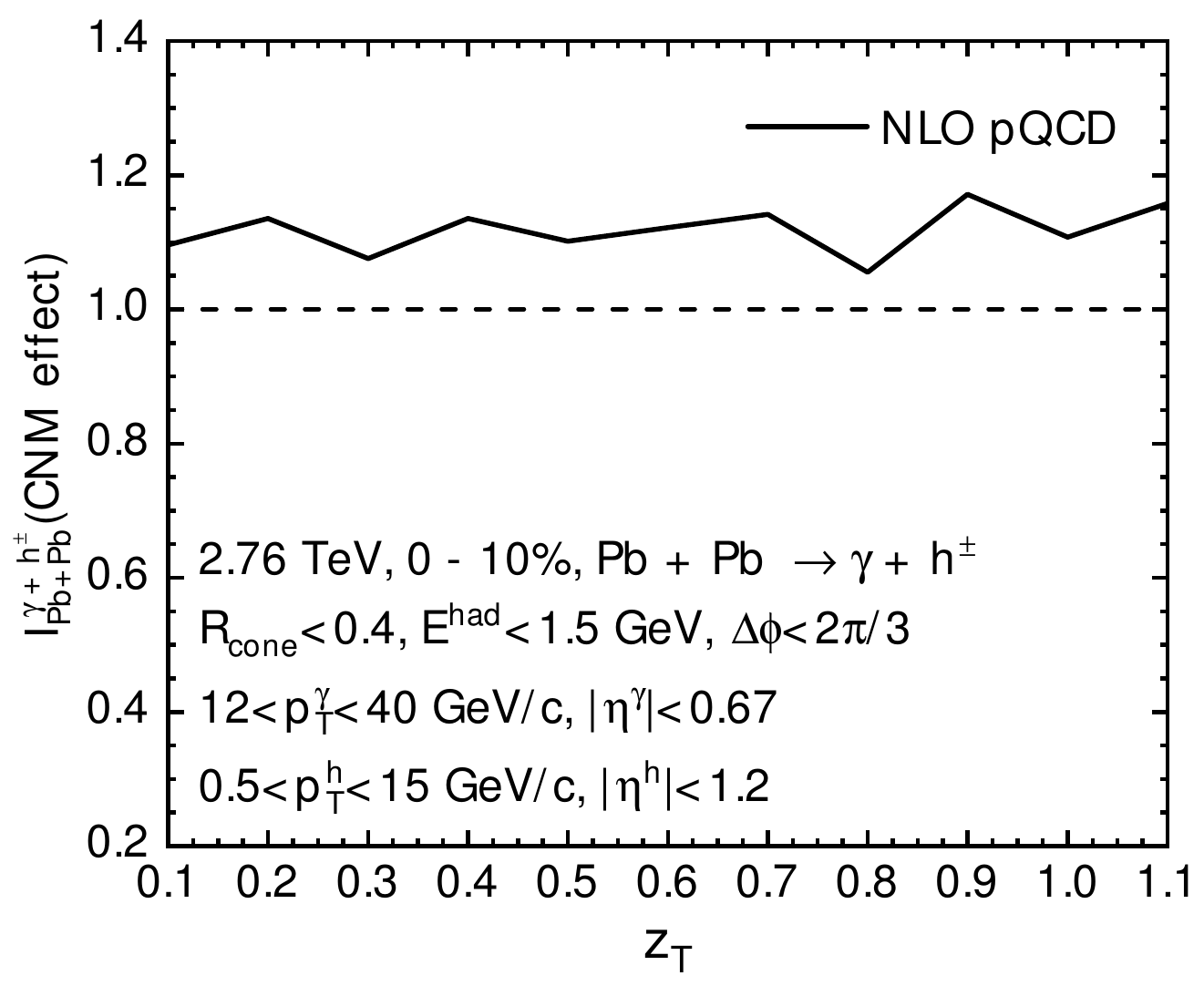}
\includegraphics[width=0.32\textwidth]{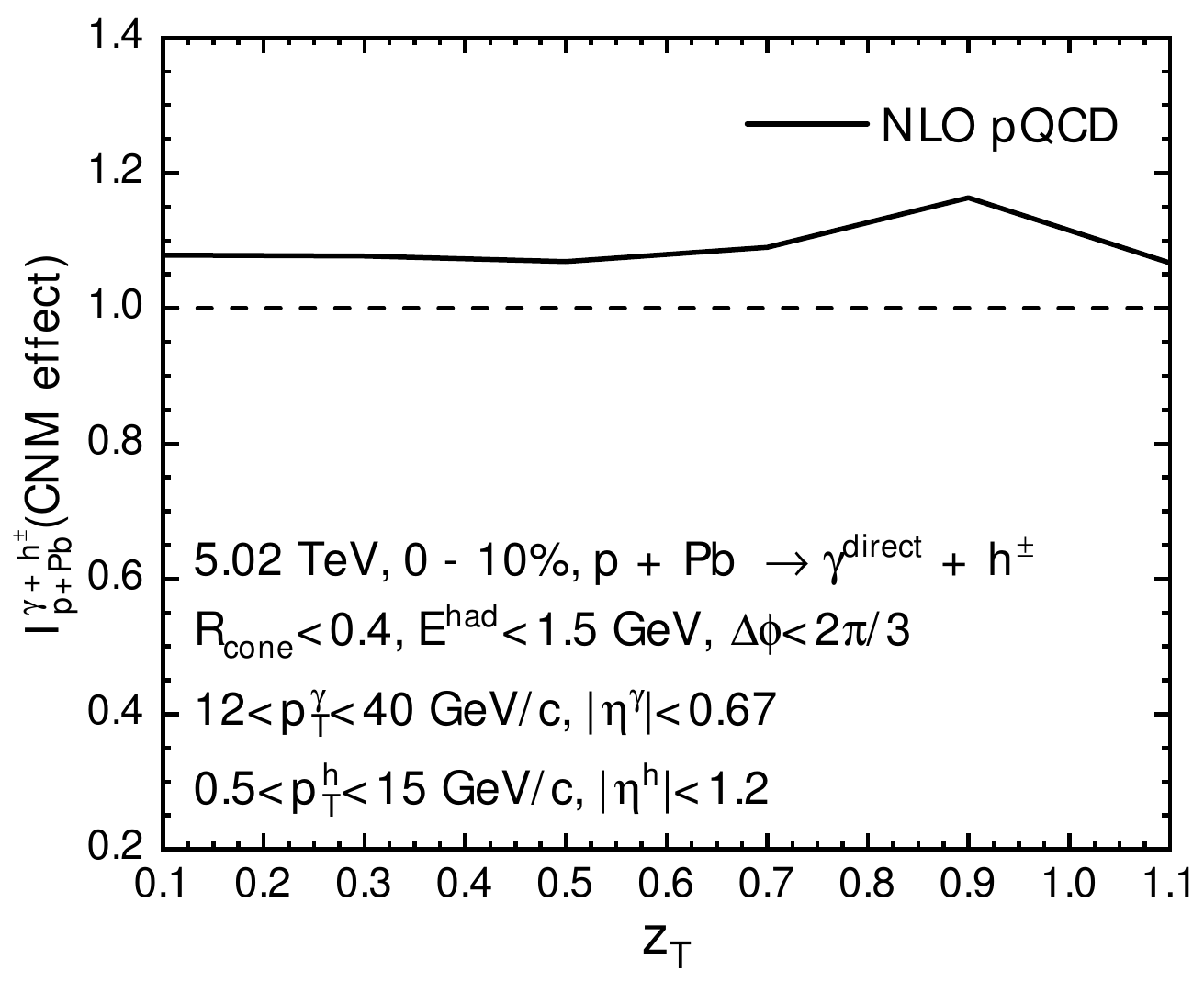}
\end{center}
\vspace{-5mm}
\caption[*]{The modification factors $I_{AA}^{\gamma h}$ or $I_{pPb}^{\gamma h}$ due to CNM effect on $\gamma^{\rm dir}$-hadron spectra in Au + Au collisions at $\sqrt{s_{\rm NN}}=0.2$ TeV (left plot), in Pb + Pb collisions at $\sqrt{s_{\rm NN}}=2.76$ TeV (middle plot) and p + Pb collisions at $\sqrt{s_{\rm NN}}=5.02$ TeV (right plot) all within 0 - 10\% centrality are shown there in correspondence with Fig. \ref{fig:CNM-AuAu} and Fig. \ref{fig:CNM-pb-ppb}.}
\label{fig:IpA-AA-CNM}
\end{figure}
\begin{figure}
\begin{center}
\includegraphics[width=0.35\textwidth]{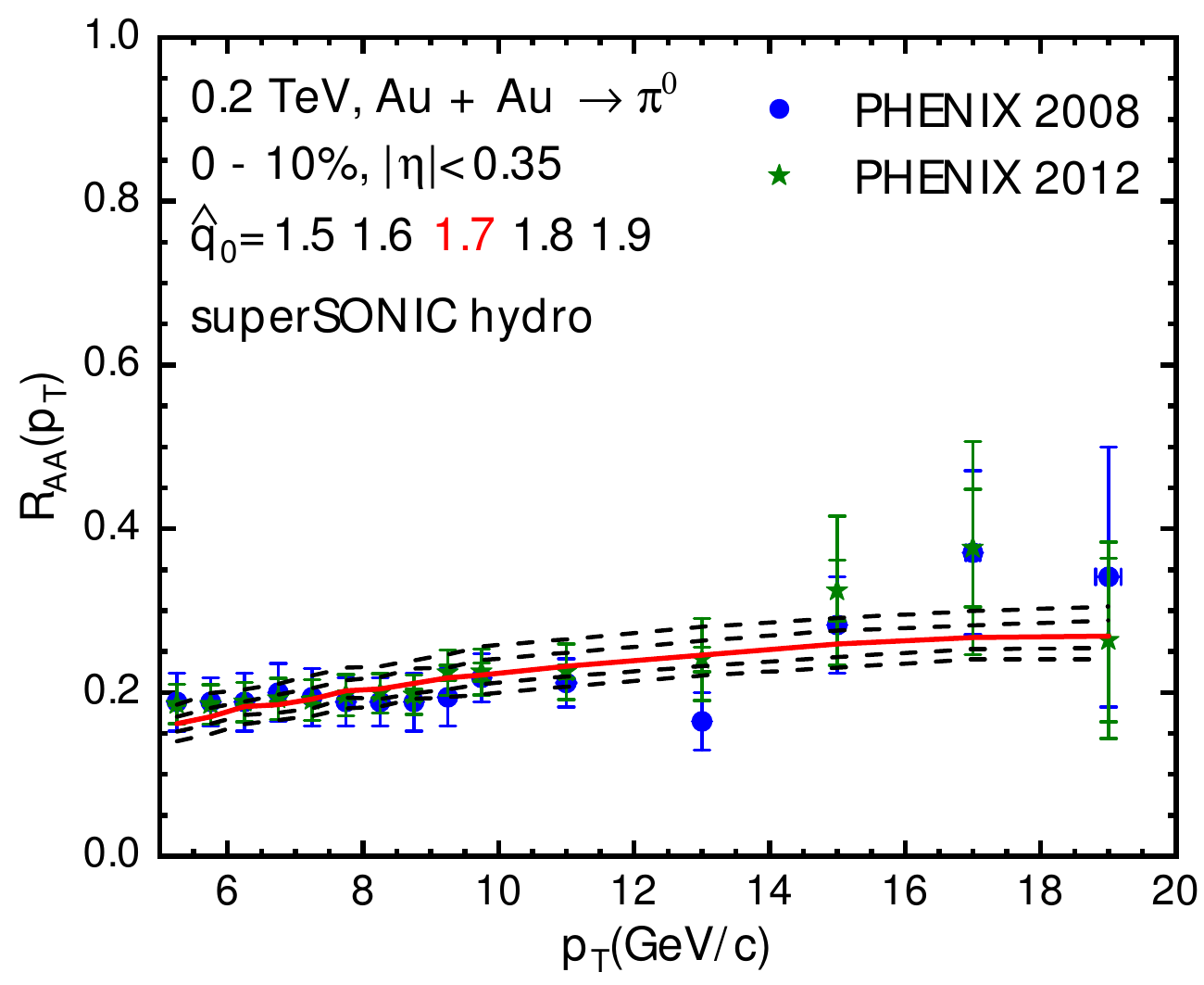}
\hspace{5mm}
\includegraphics[width=0.35\textwidth]{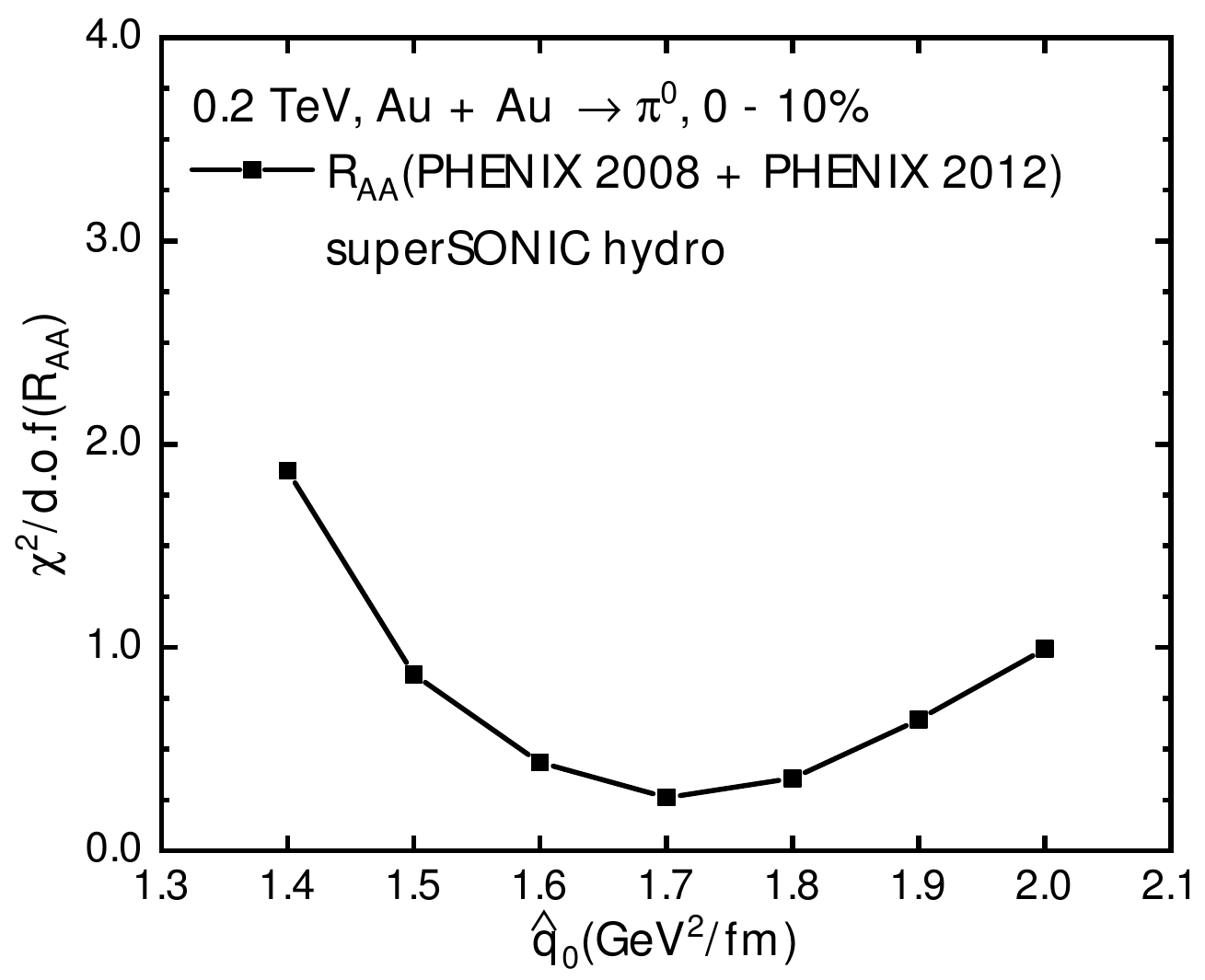}
\caption{The single hadron suppression factor (left panel) with superSONIC hydrodyanmic model in 0 - 10\% Au + Au collisions at $\sqrt{s_{\rm NN}}=0.2$ TeV compared with PHENIX \cite{Adare:2008qa,Adare:2012wg} data and the corresponding  $\chi^2/d.o.f$ of the fit as a function of the initial jet transport coefficient $\hat q_0$ (right panel).}
\end{center}
\label{fig:RHIC-super}
\end{figure}

\end{widetext}


\end{document}